\documentclass[sigconf,10pt]{acmart}
\usepackage{subfig}
\usepackage{booktabs}
\usepackage{graphicx}
\usepackage{balance}
\usepackage{algorithm}
\usepackage{algorithmicx}
\usepackage{algpseudocode}
\usepackage{caption}
\usepackage{bbm}
\copyrightyear{2019} 
\acmYear{2019} 
\setcopyright{acmlicensed}
\acmConference[SIGMOD '19]{2019 International Conference on Management of Data}{June 30-July 5, 2019}{Amsterdam, Netherlands}
\acmBooktitle{2019 International Conference on Management of Data (SIGMOD '19), June 30-July 5, 2019, Amsterdam, Netherlands}
\acmPrice{15.00}
\acmDOI{10.1145/3299869.3319857}
\acmISBN{978-1-4503-5643-5/19/06}

\newtheorem{exmp}{Example}[section]
\begin{document}
\title{Top-{\it k} Queries over Digital Traces}

\author{Yifan Li}
\email{yifanli@eecs.yorku.ca}
\affiliation{%
  \institution{York University}
}
\author{Xiaohui Yu}
\email{xhyu@yorku.ca}
\affiliation{%
  \institution{York University}
}
\author{Nick Koudas}
\email{koudas@cs.toronto.edu}
\affiliation{%
  \institution{University of Toronto}
}
\setlength\belowcaptionskip{-15pt}
\begin{abstract}
Recent advances in social and mobile technology have enabled an abundance of digital traces (in the form of mobile check-ins, association of mobile devices to specific WiFi hotspots, etc.) revealing the physical presence history of diverse sets of entities (e.g., humans, devices, and vehicles). One challenging yet important task is to identify $k$ entities that are most closely associated with a given query entity based on their digital traces. We propose a suite of indexing techniques and algorithms to enable fast query processing for this problem at scale. We first define a generic family of functions measuring the association between entities, and then propose algorithms to transform digital traces into a lower-dimensional space for more efficient computation. We subsequently design a hierarchical indexing structure to organize entities in a way that closely associated entities tend to appear together. We then develop algorithms to process top-$k$ queries utilizing the index. We theoretically analyze the pruning effectiveness of the proposed methods based on a mobility model which we propose and validate in real life situations. Finally, we conduct extensive experiments on both synthetic and real datasets at scale, evaluating the performance of our techniques both analytically and experimentally, confirming the effectiveness and superiority of our approach over other applicable approaches across a variety of parameter settings and datasets.
\end{abstract}
\maketitle
\section{Introduction}
The prevalence of mobile devices, social media, ambient wireless connectivity, and associated positioning technologies have made it possible to record digital traces at an unprecedented rate. Such traces correspond to location sharing through social apps,
handshaking with WiFi hot-spots (recording wireless chip MAC address or other device characteristics of a device in proximity of a WiFi network \footnote{The WiFi protocol reveals to the access points the MAC address of any WiFi enabled device in the vicinity of the network, even if the device is not connected to the network.}) and cellular stations via a mobile device and many other passive/active location capturing scenarios, giving rise to an abundance of {\em digital traces}. Such traces reveal the presence history of diverse sets of {\em entities} depending on the application
and include humans, devices, etc. At a high level, any digital trace takes the form of a tuple, $(entity, location, timestamp)$,
recording that an entity (e.g., a person) was present at a physical location (e.g.,  a restaurant) for the indicated timestamp.
Typically location corresponds to physical locations which exhibit a hierarchical structure that is known a priori (e.g., city - district - street - building), and the timestamp is discretized to a tunable atomic unit such as an hour or a minute, depending on the application. For example, the tuple $(Tom, W\ London, 10\ a.m.)$ represents the fact that Tom was at the W London hotel during the hour starting at 10 a.m.

A challenging task is to identify the entities that are closely associated with a given query entity utilizing their digital traces.
Intuitively two entities are associated if there exists a large overlap on their digital traces. Numerous definitions for what constitutes an overlap
are possible; for example, a large overlap in the locations followed by an overlap (proximity) of associated timestamps. Thus, if two entities were present at W London at 10 a.m., they are associated. Similarly if two entities are present at W London, one at 10 a.m. and the other at 11 a.m., they are still associated but possibly less so than the previous two entities. Alternatively, one may take into account the spatial proximity of locations to define association in addition to timestamps. Thus, if two entities are present in the same postal code at the same time, they are associated as well, but probably less so than two entities appearing at the same specific location, say a restaurant, at the same time. It is evident that there are numerous ways to define association of entities given their digital traces, which are probably application dependent. As such, we adopt a generic approach and define a class of functions that have generic properties to quantify association. All our subsequent developments in this work hold for this generic class of functions sharing such properties (see Section \ref{prob-def}).

Given a suitable function to quantify association we are interested to identify the top-$k$ associated entities to a given query entity. Supporting efficient processing of such queries over a large volume of digital traces enables a variety of applications. For example, this assists law enforcement authorities to identify individuals closely related to a person of interest. This research is motivated by our work with authorities enabling post crime investigation utilizing location data collected from mobile devices. Such information is crucial to prove the joint presence of suspects in crime scenes and also their association before and after the crime. For this specific reason, the main interest is to assess association across large sets of digital traces, corroborating the association before and after specific events. For example, in our ongoing work with a large national telecommunications provider in this problem, we will present results involving 30M individual devices with an average of 650K detections by WiFi hotspots each; in addition, each device is present on average at 500 locations during the time ranges of interest for which queries are required. In a different context, the techniques developed herein enable marketers to identify groups of individuals with related behavior in the physical world for more effective advertising. As an example, marketers may utilize associated behavior inside a shopping center (as reported by triangulated WiFi signals of mobile devices) to identify families or couples who are of prime interest for specific types of location-based marketing. Once again association across a large collection of traces enforces a closer bond between the entities involved as opposed to chance encounters.

In the target applications we  engage with, the number of entities is in the multiple millions while the number of digital traces is in the billions. As such, techniques that compare the query entity to all other entities are inefficient.

Aiming to provide fast query response times, we propose a suite of indexing structures and algorithms for this problem. At a high level, we consider entities as points in a high-dimensional space with $(location, timestamp)$ pairs of each entity corresponding to a dimension. The basic idea of our approach consists of two parts: (1) transforming an entity's digital traces into a lower-dimensional space for more efficient computation; this lower-dimensional representation also allows the ordering of entities along each dimension, making it possible to build an index structure; (2) constructing an index that groups the entities in a hierarchical fashion using this lower-dimensional representation, so that associated entities tend to appear in the same group, enabling effective pruning.

We utilize MinHash techniques \cite{broder1997resemblance} to compute a signature for each entity. When a spatial hierarchy on locations is present
we do so at each level of the spatial hierarchy, resulting in a list of signatures for each entity. The size of each signature depends on the number of hash functions utilized and can be considered as the dimensionality of this lower-dimensional space. 
The list of signatures for each entity are subsequently indexed with a tree structure. The guiding principal of this index is to group the entities based on their signatures at each level such that for a given query entity, only a small portion of the branches in the tree have to be explored; the remaining branches are guaranteed not to contain the top-$k$ results and can thus be safely discarded. This is made possible by assessing a signature for each group of entities at a tree node, which serves as the basis for estimating bounds on the association between the query entity and the entities in the subtree rooted at this node. The index naturally supports incremental updates. We then develop algorithms to process top-$k$ queries using the index.

We also develop and present a model for mobility which we validate against real data at scale. Utilizing this model, we subsequently present a thorough analysis of the pruning effectiveness of the proposed method. Our results reveal that the proposed technique has strong pruning capabilities, limiting the scope of search to only a small portion of all available entities. We also validate our model for pruning effectiveness against real mobility traces and demonstrate its accuracy both analytically and experimentally.

Experiments are conducted on both synthetic and real datasets at scale to (1) compare the performance of the proposed method against that of baseline methods, and (2) conduct a sensitivity analysis of the proposed method with respect to varying parameters of interest (e.g., number of hash functions, data characteristics). Our results demonstrate orders of magnitude performance improvement over other applicable approaches.

In summary, in this paper we make the following contributions:
\begin{itemize}
\item Motivated by real-life applications with telecommunications providers, we formally define the problem of Top-$k$ query processing over digital traces. To the best of our knowledge, our work is the first to address this important problem.
\item We develop a suite of novel data transformation and indexing techniques as well as the corresponding search methodologies, which demonstrate strong pruning capabilities, allowing us to focus the search only on a small portion of the space.
\item We analytically and experimentally quantify the pruning effectiveness of our methods utilizing models of human mobility patterns.
\item We perform extensive experiments on both real and synthetic data at scale to thoroughly study the performance of the proposed method, confirming its effectiveness and superiority over other approaches across a variety of settings.
\end{itemize}
The rest of the paper is organized as follows. Section 2 formally defines the problem of top-$k$ query over digital traces and other assist terms. Section 3 describes the approach, including the data transformation principle, the data organization technique, and the complexity of indexing. In Section 4, we prove the early termination condition of the proposed approach and give the search algorithm. In Section 5, we analytically quantify the pruning effectiveness of the approach. In Section 6, we present the experiment results across a variety of settings. Section 7 provides an overview of related work, and Section 8 concludes this paper.

\section{Preliminaries}
In this section, we define the terms that are required for the subsequent discussion, and formally define the problem of top-$k$ query processing over digital traces.
\subsection{Terminology}\label{sec:term}
The locations we consider are spatial and thus exhibit a hierarchical structure (e.g., city - district - street - building). We assume that a description of the hierarchical structure of locations is available via a tree structure (referred to as {\em sp-index}) that organizes locations from coarsest to finest. Nodes in this tree are referred to as {\em spatial units}. We assume (as per previous work \cite{wang2016crime}) that spatial units remain unchanged over extended times periods, and thus the sp-index can be considered fixed for the period of interest.

Without loss of generality, we assume that the spatial units at the same level of the sp-index are non-overlapping. We label the levels of spatial units from $1$ (for the root of the tree) to $m$ (for the lowest level in the tree). For a spatial unit $l$, we use pat($l$) to denote the parent unit of $l$ on the sp-index.

At the lowest level of the tree are \textit{base spatial units}, the atomic locations in digital traces in which entities can be present. Examples of a base spatial unit include a supermarket, a restaurant, etc. All base spatial units form a set $\mathcal{L} $.

We assume that timestamp is discretized in base temporal units (e.g., hour). The combination of a base temporal unit and a base spatial unit is referred to as a {\em spatial-temporal cell} (or ST-cell). We use the associated base temporal unit and base spatial unit to denote an ST-cell, e.g., $t_1l_1$. An ST-cell is the atomic unit where entities can be present. All possible such combinations form an ST-cell set $\mathcal{S}$.

We enhance the notion of a digital trace associated with entity $e$ to make it suitable for a multilevel sp-index.

\begin{definition}[Presence Instance]
A presence instance (PI) $p$ of an entity is characterized by a five attribute tuple,
$p = (e,tid,level,path,pd)$, where
\begin{itemize}
\item $e$ is the associated entity to $p$,
\item $tid$ is the id of the sp-index where $p$ belongs ($tid$ is necessary when multiple sp-index trees exist),
\item $level$ is the level in the sp-index where $p$ exists,
\item $path = [node_{1}, node_{2},\cdots, node_{level}]$ is the list of nodes in the sp-index on the path from the root to the node that reflects the location associated with $p$, and
\item $pd$ is a continuous time period associated with $p$; it is in the format $[start\ time, end\ time]$.
\end{itemize}
\end{definition}

Typically $start\ time$ is the same as $timestamp$. In some applications, such as WiFi proximity sensing of MAC addresses, $end\ time$ is obtained by the time of the last probe of the device MAC address to the WiFi network. In some other applications, such as social media check-ins, the end time is estimated based on the average time individuals spend in the venue (obtained from services such as Google Maps).
\begin{definition}[Digital Trace]
The set of PIs associated with entity $e$ forms the digital trace of $e$, $\mathcal{P}_e$.
\end{definition}
The overlap between the digital traces of two entities, {\em Adjoint Presence Instance}, is defined as follows:
\begin{definition}[Adjoint Presence Instance]
Given two PIs, $p_a = (e_a, tid_a, level_a, path_a, pd_a)$, $p_b = (e_b, tid_b, level_b,path_b,\\ pd_b)$, if $tid_a=tid_b$, $pd_a\cap pd_b\neq \emptyset$, then $e_a$ and $e_b$ form an adjoint presence instance (AjPI) $p_{ab}=(\{e_a,e_b\},tid_{ab},level_{ab},\\ path_{ab},pd_{ab})$, where:
\begin{itemize}
\item $tid_{ab} = tid_a=tid_b$,
\item $level_{ab} = \vert path_{ab}\vert$, denoting the finest level of the AjPI, which is equal to the number of common ancestors in the sp-index,
\item $path_{ab} = path_a \cap path_b$, which is the set of common ancestors of the two PIs, and
\item $pd_{ab} = pd_a\cap pd_b$, the intersection of the two time periods.
\end{itemize}
\end{definition}
Each pair of entities, say $e_a$ and $e_b$, own zero or more AjPIs, forming set $\mathcal{P}_{ab}$. The definition can be naturally extended to adjoint presence instances of multiple entities.

An AjPI specifies a spatio-temporal co-occurrence of two entities, and thus reveals a potential association between the entities. Such an association
is defined as a function of the corresponding presence instances and adjoint presence instances of each entity pair, as outlined next.

\subsection{Problem definition}
\label{prob-def}
One important but challenging task is to discover all entities that are closely associated with a given entity. Since there may exist many ways to quantify association, we define a generic class of scoring functions that share some commonly desired properties. While the particular choice of the function varies depending on the application, our approach would apply as long as the function exhibits those generic properties.

For an AjPI $p_{ab}$, the scoring function $f(p_{ab})$ has the following properties:
\begin{itemize}
  \item The range of $f\in [0,1]$,
  \item $\forall e_c$, $f(p_{ab})\geq f(p_{ac})$\ if\ $p_{ab}.pd.length \geq p_{ac}.pd.length\\ \wedge p_{ab}.level \geq p_{ac}.level$.
\end{itemize}
The first property ensures that the score is properly normalized; the second property gives AjPIs at finer spatial units and for longer durations a higher score. These properties capture the intuition that the association between two entities is higher when corresponding digital traces match closely at locations (i.e.,  appear at finer levels of the sp-index, say at the same restaurant vs. in the same city) and their temporal co-occurrence is longer.

Let $\mathcal{P}_{ab}$ be the set of AjPIs formed by $e_a$ and $e_b$. The overall score for this set is defined as
\begin{equation}
  F(\mathcal{P}_{ab}) = \sum\limits_{p_{ab}\in \mathcal{P}_{ab}}f(p_{ab}),
\end{equation}
which has to be further normalized to take into consideration the individual behaviors of $e_a$ and $e_b$. It is evident that the AjPI to an entity with many PIs is less interesting than that with an entity having few PIs. Therefore, we define a scoring function for individual PI $p_a$, which is considered as a special case of the AjPI score, i.e., $f(p_a) = f({p}_{aa})$. The score for the set of PI $\mathcal{P}_a$ is:
\begin{equation}
  F({\mathcal{P}_{a}}) = \sum\limits_{p_{a}\in \mathcal{P}_{a}}f(p_{a})
\end{equation}
Clearly, $\forall e_a, \forall e_b, F(\mathcal{P}_{a})\geq F(\mathcal{P}_{ab}), F(\mathcal{P}_{b})\geq F(\mathcal{P}_{ab})$.

Intuitively, closely associated entities are those who have a large presence instance overlap, i.e., more adjoint presence instances and less total presence instances for either entity. Thus we define the association degree between two entities $e_a$ and $e_b$ as
\begin{equation}\label{eq:d}
d(e_a,e_b) = G(F(\mathcal{P}_{ab}),F(\mathcal{P}_{a}),F(\mathcal{P}_{b}))
\end{equation}
where $G$ can be any function satisfying the following constraints:
\begin{itemize}
\item $d(e_a,e_b)$ has range $[0,1]$,
\item $\forall e_c$, $d(e_a,e_b)\geq d(e_a,e_c)$ if $F(\mathcal{P}_{b})\leq F(\mathcal{P}_{c}) \wedge F(\mathcal{P}_{ab})\geq F(\mathcal{P}_{ac})$,
\item $\forall e_c$, $d(e_a,e_b)\geq d(e_a,e_c)$ if $(\mathcal{P}_b-\mathcal{P}_c)\neq \emptyset \wedge (\mathcal{P}_b-\mathcal{P}_c)\subseteq \mathcal{P}_a$.
\end{itemize}
We let the associate degree be a generic function instead of a particular measure (e.g., Jaccard Distance), as the most suitable measure may vary in different application scenarios. A generic approach allows our Top-$k$ algorithm to utilize the measure that makes the most sense in each case as long as it follows set properties. Therefore, all subsequent discussions of Top-$k$ query processing are for measures satisfying the constraints of $G$. We provide a recommended form of $G$ in Section \ref{settings}, where We also demonstrate that such a form simulates other widely-adopted measures accurately.

Let $\mathcal{E}$ be the set of all entities. The problem of identifying the $k$ most associated entities (entities with the highest association degree) to a given query entity is defined as:
\begin{definition}[Top-$k$ Query over Digital Traces]
Given a query entity $e_p$ and association degree measure $d$, the top-$k$ query over digital traces is to return the set of entities $\mathcal{Q}_k$ such that $\mathcal{Q}_k\subseteq \mathcal{E}-\{e_p\}$, $\vert \mathcal{Q}_k\vert=k$ and $\forall e_q\in \mathcal{Q}_k,\ \forall e_t \in (\mathcal{E}-\{e_p\}-\mathcal{Q}_k),\ d(e_p,e_q)\geq d(e_p,e_t) $, where $1\leq k< \vert \mathcal{E}\vert$.
\end{definition}
\section{Our Approach}
\label{sec:approach}
A brute-force approach to answer top-$k$ queries involves computing the association degree between the query entity and all other entities. Clearly, the cost can be prohibitive, as the number of entities are often in the millions and the number of digital traces in billions in our target applications.
As such, we introduce a data structure, called the {\em MinSigTree}, that indexes entities based on their presence instances, facilitating efficient pruning of entities to be examined during the search for top-$k$ answers.

As a high-level overview, we first organize the PIs of each entity as a sequence of ST-cell sets. Then we construct a list of {\em signatures} for each entity which can be considered as summaries of the entity's PIs. Subsequently we construct the MinSigTree that groups closely associated entities together based on their signatures.

\subsection{Data representation}
\label{sec:representation}
Real-world digital traces in their raw format may require pre-processing. For instance, we may need to conduct time-zone normalization (e.g., all time-stamps normalized to GMT), build the sp-index from the longitude and latitude coordinates of places using maps (e.g., Open Street Maps) and other information (e.g., community area boundaries), and align data from different sources with varying sampling frequencies. After pre-processing, the next step is to organize the data by entity so that the presence instances of an entity at each sp-index level and the resulting association degree between entity pairs can be computed efficiently.

We build a sequence of ST-cell sets for each entity, where the length of the sequence equals the height of the sp-index, $m$. The sequence of sets for entity $e_a$ is denoted as $seq_a$, and the $i$-th set in $seq_a$, $i\in[1,m]$, corresponding to the level $i$ of the sp-index, is denoted as $seq_a^i$.

$seq_a^m$, for the lowest level of the sp-index, can be obtained directly from $e_a$'s digital trace, i.e., for an ST-cell $s$, $s\in seq_a^m$ iff $e_a$ is present at $s$. For other levels, ST-cell set $seq_a^i$, $i\in[1,m)$ is built from set $seq_a^{i+1}$. For an ST-cell $s=t_zl_x$, $s\in seq_a^i$ iff $\exists s'=t_zl_y$, s.t. $s'\in seq_a^{i+1}$ and $l_x=$pat$(l_y)$.
\begin{exmp}
\label{exmp:seq}
Let $L_1$, $L_2$, $L_3$, and $L_4$ be four base spatial units, and pat($L_1$)=pat($L_2$)=$L_5$, pat($L_3$)=pat($L_4$)=$L_6$, $m=2$. Assume that entity $e$ has presence in base spatial unit $L_1$ at time $T_1$, and $L_3$ at time $T_2$, then $seq_e^2=\{T_1L_1,T_2L_3\}$. Since $T_1L_1\in seq_e^2$ and $L_5=$pat$(L_1)$, $T_1L_5\in seq_e^1$, similarly, $T_2L_6\in seq_e^1$. Finally we have $seq_e^1=\{T_1L_5,T_2L_6\}$.
\end{exmp}

The ST-cell set sequence not only records the PIs of a single entity at any level of the sp-index, but reflects the AjPI between entities as well. If entities $e_a$ and $e_b$ form AjPIs at level $i$, then $seq_a^i\cap seq_b^i\neq \emptyset$.

\subsection{Data organization}
Although  ST-cell set sequences facilitate the direct retrieval of PIs of each entity at any level, a brute-force approach would have to explore the whole search space of all entities to identify the top-$k$ answers, which is still too expensive.
We thus propose to group entities based on their common ST-cells to allow efficient pruning of the search space.  Note that the number of ST-cells in which an entity is present could vary vastly from entity to entity (e.g., one short occurrence vs. frequent and prolonged visits to multiple locations). If we consider each ST-cell as a dimension, conceptually all entities can be considered as bit vectors in a very high-dimensional space where each bit indicates whether that entity is present in the ST-cell. However, if they are physically treated as such, the  storage and computation cost can be prohibitive when the number of ST-cells is large.  To allow more effective indexing, we employ a family of hash functions to map ST-cell sets into a lower-dimensional space. This is achieved by assigning each entity a \textit{signature} at each level, with each value in the signature acting as a summary of the entity's PIs, and then grouping entities by their signatures. 
\subsubsection{Signature}
\label{sec:signature}
We use $n_h$ hash functions to map an ST-cell set into a vector in a $n_h$-dimensional space, where each element of the vector is a hash value. Since each entity is associated with an ST-cell set sequence of length $m$, for an arbitrary entity $e_a$ ,we obtain $m$ vectors, which form a list of signatures, $sig_a$, and we use $sig_a^i$ to denote the $i$-th signature in $sig_a$ (corresponding to level $i$ in the sp-index), and $sig_a^i[u]$ to denote the $u$-{th} hash value in $sig_a^i$, where $u\in [1,n_h]$.

The way we compute signatures for each entity is similar to that for MinHash \cite{broder1997resemblance}. A hash function $h_u$ maps each ST-cell to a value in the range $[0,\mathcal{\vert S\vert}-1]$. The $u$-th value in the signature $sig_a^i$ corresponds to the minimal hash value produced by $h_u$ across all ST-cells in $seq_a^i$, i.e., $sig_a^i[u] = \bot^i_u=\min_{s\in seq_a^i}\{h_u(s)\}$. 

The hash functions employed above should satisfy that, for ST-cell $s=t_zl_x$ and $s'=t_zl_y$, if $l_x=$pat$(l_y)$, then $h_u(s) \leq h_u(s')$. Let $\mathcal{C}_x$ be the child spatial unit set of $l_x$, the above constraint is satisfied by assigning $h_u(t_zl_x)=\min_{l_c\in \mathcal{C}_x}\{h_u(t_zl_c)\}$. The constraint guarantees the following property which makes signatures at different levels comparable:

\begin{theorem}
\label{theo:level}
For any entity $e\in \mathcal{E}$, $sig_e^{i}[u] \leq sig_e^{i+1}[u]$ always holds.
\end{theorem}
The proof follows from above constraint and is omitted for brevity.

\begin{exmp}
\label{exmp:signature}
Consider the following hash table:

\begin{center}
\begin{tabular}{r c c c c c c c c}
\hline
  & $T_1L_1$ & $T_2L_1$ & $T_1L_2$ & $T_2L_2$ & $T_1L_3$ & $T_2L_3$ & $T_1L_4$ & $T_2L_4$ \\ 
 \hline
 $h_1$ & 2 & 8 & 5 & 1 & 4 & 6 & 7 & 3\\ 
 \hline
 $h_2$ & 8 & 3 & 6 & 5 & 4 & 1 & 2 & 7\\ 
 \hline
\end{tabular}
\end{center}
Assume that the four base spatial units follow relations indicated in Example \ref{exmp:seq}. Let $e_a$, $e_b$, $e_c$ and $e_d$ be four entities with the following ST-cell set sequence:
\begin{center}
 \begin{tabular}{c c}  
 \hline
  $e_a$ & $\langle\{T_1L_5,T_2L_5\},\{T_1L_2,T_2L_1\}\rangle$\\ 
 \hline
  $e_b$ & $\langle\{T_1L_5,T_2L_5\},\{T_1L_1,T_2L_2\}\rangle$\\ 
  \hline
  $e_c$ & $\langle\{T_1L_6,T_2L_5\},\{T_1L_3,T_2L_1\}\rangle$\\ 
  \hline
  $e_d$ & $\langle\{T_1L_6,T_2L_6\},\{T_1L_4,T_2L_4\}\rangle$\\ 
 \hline
\end{tabular}
\end{center}
We first build $sig_a^2$ given $seq_a^2 = \{T_1L_2,T_2L_1\}$. Since $h_1(T_1L_2)=5$, $h_1(T_2L_1)=8$, we have $sig_a^2[1]=5$; similarly, since $h_2(T_1L_2)=6$, $h_2(T_2L_1)=3$, we have $sig_a^2[2]=3$. Therefore, $sig_a^2 = \langle5,3\rangle$. Subsequently, we build $sig_a^1$ given $seq_a^1 = \{T_1L_5,T_2L_5\}$. Since $L_5 = pat(L_1)=pat(L_2)$, $h_1(T_1L_5)=\min\{h_1(T_1L_1),h_1(T_1L_2)\}=2$; similarly we have $h_1(T_2L_5)=1$, $h_2(T_1L_5)=6$, $h_2(T_2L_5)=3$. Therefore, $sig_a^1 = \langle 1,3\rangle$. We build signatures for all entities and finally obtain the following signature table:
\begin{center}
 \begin{tabular}{c c}  
 \hline
  $e_a$ & $\langle\langle 1,3\rangle, \langle 5,3\rangle \rangle$\\ 
 \hline
  $e_b$ & $\langle \langle 1,3\rangle, \langle 1,5\rangle \rangle$\\ 
  \hline
  $e_c$ & $\langle \langle 1,2\rangle, \langle 4,3\rangle \rangle$\\ 
  \hline
  $e_d$ & $\langle \langle 3,1\rangle, \langle 3,7\rangle \rangle$\\ 
 \hline
\end{tabular}
\end{center}
\end{exmp}
As each value in a signature is obtained by hashing all ST-cells in the corresponding set to a certain domain, it can be considered as a summary of the ST-cell set. Hash values $sig_a^i$ enables to determine certain facts regarding the ST-cells contained in the set $seq_a^i$.
\begin{theorem}
\label{th:exclusive}
For signature $sig_a^i$ $(i\in[1,m])$ and an ST-cell $s$, if $\exists u\in [1,n_h]$ s.t. $sig_a^i[u]>h_u(s)$, then $s\notin seq_a^m$.
\end{theorem}

\begin{proof}
If $s\in seq_a^m$, then $sig_a^m[u]\leq h_u(s)$. From Theorem \ref{theo:level} we know that $sig_a^i[u]\leq sig_a^m[u]$, and thus $sig_a^i[u]\leq h_u(s)$, which contradicts the condition.
\end{proof}

Via Theorem \ref{th:exclusive}, for a given signature $sig$, we can obtain a \textit{pruned set} of ST-cells such that entities bearing $sig$ are guaranteed not to have presence in those ST-cells. We use $\mathcal{PS}_a^i$ to denote the pruned set based on signature $sig_a^i$. This property will be explored in pruning the search space while computing the top-$k$ answers.
\subsubsection{MinSigTree}
\label{sec:minsigtree}
We design MinSigTree, an $m$-level tree structure, which groups entities sharing similar signatures together.
Each node in the MinSigTree has at most $n_h$ child nodes (with $n_h$ being the number of hash functions used while computing signatures), each leaf node contains a set of entities, and each entity is contained in a single leaf node. If node $N$ contains entity $e_a$, we consider all ancestor nodes of $N$ to conceptually contain $e_a$ as well to ease notation (but no physical storage is involved). For node $N$ containing entity set $\mathcal{E}_N$, we compute a group-level signature $SIG_N$ summarizing the PIs of all entities in $\mathcal{E}_N$.

Assuming that there is a virtual root node (at level 0), we use Algorithm 1 to build the MinSigTree.
\begin{algorithm}[htb]
\caption{Building MinSigTree}
\label{alg:Framwork}
\begin{algorithmic}[1]
\Require
Entity set $\mathcal{E}$, signatures of all entities
\Ensure
MinSigTree
\State \textbf{Initialization:} MinSigTree root to contain all entities; root enqueued to priority queue $Q$;
\For {$N$ : $Q$}
    \State $\mathcal{G}$ = sets of entities in $N$ grouped by routing index;
    \For {$g$ : $\mathcal{G}$}
        \State $u = $ routing index of $g$;
        \State $\mathcal{E}_g = $ entities contained in $g$;
        \State $SIG_g = $ group-level signature of $\mathcal{E}_g$;
        \State $N_u = $ node($u$, $SIG_g[u]$, $\mathcal{E}_g$);
        \State $N$.addChild($N_u$);
        \If {$i\neq m$}
            \State  enqueue $N_u$to $Q$;
        \Else
            \State insert $\mathcal{E}_g$ to $N_u$;
        \EndIf
    \EndFor
\EndFor
\State \Return MinSigTree;
\end{algorithmic}
\end{algorithm}

As Step 1, we fetch the level 1 signature of every entity, $(sig_1^1, sig_2^1, \cdots, sig_{\vert\mathcal{E}\vert}^1)$, and divide these signatures into $n_h$ groups. This is done in a way such that entity $e_a$ is routed to the $u$-{th} group, if $\forall v\in[1, n_h] (v\neq u), sig_a^1[u]\geq sig_a^1[v]$, i.e., $u$ is the position of the maximal hash value in $sig_a^1$ (ties are broken arbitrarily). We call $u$ the \textit{routing index} of the $u$-{th} group (Line 3).

Step 2 involves computing a group-level signature for each node (Lines 5 - 7). Assume that node $N_u$ contains entity set $\mathcal{E}_{N_u}$. Then the signature of $N_u$, $SIG_{N_u}$, can be computed by $SIG_{N_u}[v]=\min_{e\in \mathcal{E}_{N_u}}\{sig_e^1[v]\}$, where $v\in[1, n_h]$. The newly created nodes are then inserted as the children of the root (Lines 8 - 9).

The second step computes a group-level signature for each node in a way that any hash value in $SIG_{N_u}$ is no greater than the corresponding hash values in the signatures of entities in $\mathcal{E}_{N_u}$. With signatures computed this way, we can obtain a group-level pruned set,
\begin{equation}
\mathcal{PS}_{N_u}=\bigcap\limits_{e\in \mathcal{E}_{N_u}}\mathcal{PS}_{e}^1.
\end{equation}
All entities in $\mathcal{E}_{N_u}$ are guaranteed not to have presence in the ST-cells contained in $\mathcal{PS}_{N_u}$. Note that there is no need to store the pruned set of each node, as it can be inferred from the group-level signature.

In practice, however, storing the entire signature of a node imposes space overhead. It is evident from the grouping strategy that given a group-level signature $SIG_N$ with routing index $u$, $\forall v\in[1, n_h] (v\neq u)$, $SIG_N[u]\gg SIG_N[v]$. From Theorem \ref{th:exclusive} it follows that the pruned set of a signature is mainly decided by the large hash values in the signature. Thus one can materialize $SIG_N[u]$ only, instead of $SIG_N$. This greatly reduces storage costs at the expense of pruning effectiveness. We explore this further in Section \ref{sec:earlytermination}.

Consider the signature table in Example \ref{exmp:signature}. We fetch all level $1$ signatures and group entities accordingly. As a result, group $N_1=\{e_d\}$ with routing index $1$, $N_2=\{e_a, e_b, e_c\}$ with routing index $2$, and $SIG_{N_1}=\langle 3,1\rangle$, $SIG_{N_2}=\langle 1,2\rangle$.

The grouping principle of Step 1 is designed in a way to prevent the group-level signature from becoming too small. For example, if $e_c$ and $e_d$ were to be grouped together, the group-level signature would be $\langle 1,1\rangle$, which would not be greater than any hash values and the pruned set would thus be empty.

Now we have grouped entities at the first level of the MinSigTree based on the level 1 signatures of all entities. However, the level 1 signatures reveal only the PI patterns at the highest/coarsest sp-index level.  Intuitively, entities belonging to different groups at level $1$ are guaranteed not to be strongly associated, but entities belonging to the same group may still have different PI patterns at a finer-level. For example, if two people both visited New York City, but one in Manhattan and the other in Brooklyn, their PIs are different in the district level. Therefore, we need to further partition the entities based on their finer-level signatures.

For node $N_u$ at level $i$, if $i\neq m$, i.e. $N_u$ is not at the leaf level, we fetch the level $(i+1)$ signatures of entities in $\mathcal{E}_{N_u}$ (Lines 10 - 11), group $\mathcal{E}_{N_u}$ by the routing indexes, compute a signature for each new group, and add these newly created nodes as children of $N_u$. We repeat this process until we reach the leaf level. If an entity belongs to node $N_f$ at the leaf level, we insert this entity to $N_f$ (Lines 12 - 13).

\begin{figure}
\centering
\includegraphics[width=0.5\columnwidth]{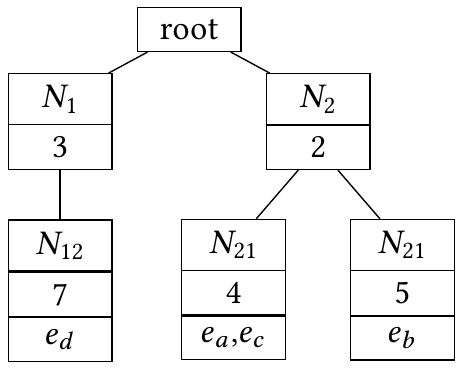}
\vspace{-3mm}
\caption{A sample MinSigTree}
\label{fig:MinSigTree}
\end{figure}

In the above example, group $N_1=\{e_d\}$, $N_2=\{e_a,e_b,e_c\}$. Since $sig_d^2[2]>sig_d^2[1]$, $e_d$ belongs to the sub-group with routing index $2$, i.e. $N_{12}=\{e_d\}$; similarly, we have $N_{21}=\{e_a,e_c\}$, $N_{22}=\{e_b\}$. Group level signatures are: $SIG_{N_{12}}=\langle 3,7\rangle$, $SIG_{N_{21}}=\langle 4,3\rangle$, $SIG_{N_{22}}=\langle 1,5\rangle$. The overall MinSigTree is given in Figure \ref{fig:MinSigTree}.

By partitioning entities recursively at each level, each group will end up containing entities that are similar at all sp-index levels and thus very likely to result in high association degrees with each other. In addition, the partitioning strategy guarantees the following property.
\begin{theorem}
\label{th:ancestor}
If $N_a$ is an ancestor node of $N_d$, then $\mathcal{PS}_{N_a}\subseteq\mathcal{PS}_{N_d}$.
\end{theorem}
The proof follows from the building process and is omitted for brevity.

The cost of index construction is analyzed in Appendix \ref{appen:cost}.
\subsubsection{Incremental update}
\label{sec:incrementalupdate}
Similar to the building process, the MinSigTree also supports incremental update. More specifically, after building the MinSigTree, we can deal with new records of entity $e$ in four steps: (1) locate the leaf, say $N_e$, containing $e$; (2) remove $e$ from $N_e$ (if $N_e$ becomes empty, safely remove $N_e$ from the MinSigTree); (3) compute new signatures of $e$; (4) insert $e$ to the new node $N_e'$ based on the new signatures. During this process, only nodes in the path from root to the leaf containing $e$ and the leaf to insert $e$ are modified, and thus the complixity of update is linear w.r.t. the height of the MinSigTree.

Bulk updates are also naturally supported. Here we mainly introduce the steps involving signature re-computation. Given a set of entities to be updated, $\mathcal{E}_u$,  we compute the signatures of $e$ for each $e\in\mathcal{E}_u$, and use $\mathcal{E}_N$ to denote the set of entities in $\mathcal{E}_u$ to be inserted to node $N$. Subsequently, we compute the group level signature of $\mathcal{E}_N$, $SIG_{\mathcal{E}_N}$, and update the signature of $N$, $SIG_N$, as $SIG_{N}:=\min \{SIG_{N}, SIG_{\mathcal{E}_N}\}$. Updating MinSigTree is discussed in detail in Section \ref{sec:indexingcost}. 

\section{Query Processing}
The MinSigTree partitions entities to groups enabling an efficient search strategy for top-$k$ query processing. We present an algorithm for top-$k$ query evaluation utilizing the proposed structure.
\subsection{Early termination}
\label{sec:earlytermination}
Given a query entity $e_q$ with ST-cell set sequence $seq_q$, the basic search strategy unfolds by computing an upper bound on the association degree between $e_q$ and each candidate node of the MinSigTree, and then progressively visit the node with the maximal upper bound until the top-$k$ answers are identified. We outline how to compute and gradually tighten the upper bound of a node in order to prune more entities and terminate the search earlier.

We use $\mathcal{S}_q$ to denote $seq_q^m$, which contains all ST-cells in which $e_q$ is present. For each node $N$ in the search path, we determine an upper bound, $UB_N$, on the association degree between $e_q$ and entities in node $N$, to decide whether to continue searching or terminate.
\begin{theorem}
Let $\mathcal{PS}_N$ be the pruned set of node $N$, and $e_v$ be an artificial entity with ST-cell set $\mathcal{S}_v=\mathcal{S}_q-\mathcal{PS}_N$. Then $UB_N=d(e_v, e_q)$.
\end{theorem}
\begin{proof}
Because $\mathcal{S}_v\subseteq \mathcal{S}_q$, we have $\mathcal{P}_v\subseteq \mathcal{P}_q$, where $\mathcal{P}_v$ and $\mathcal{P}_q$ are the PI sets of $e_v$ and $e_q$ respectively. Thus, $\mathcal{P}_{vq} = \mathcal{P}_v$, where $\mathcal{P}_{vq}$ is the set of AjPIs between $e_v$ and $e_q$.

Let $\mathcal{E}_N$ denote the set of entities contained in $N$. Since $\forall e_p\in \mathcal{E}_N$, $\mathcal{S}_p \cap \mathcal{S}_q \subseteq \mathcal{S}_v$, we have $\mathcal{P}_{pq} \subseteq \mathcal{P}_v=\mathcal{P}_{vq}$. Therefore, $F(\mathcal{P}_{pq})\leq F(\mathcal{P}_{vq})$.

$\forall e_p\in \mathcal{E}_N$, if $\mathcal{P}_v \subseteq \mathcal{P}_p$, then $F(\mathcal{P}_p)\geq F(\mathcal{P}_v)$, and thus $d(e_v,e_q)\geq d(e_p,e_q)$; otherwise,we have $(\mathcal{P}_v - \mathcal{P}_p)\neq \emptyset$ and $(\mathcal{P}_v - \mathcal{P}_p)\subseteq \mathcal{P}_q$, and thus $d(e_v,e_q)\geq d(e_p,e_q)$ according to the definition of $d$ given in Equation~(\ref{eq:d}).
\end{proof}

In practice, it is not required to compute the entire pruned set of a node; instead, it can be conducted in a more efficient way. Let $u$ be the routing index of the node $N$. For an ST-cell $s\in \mathcal{S}_q$, if $h_u(s)<SIG_N[u]$, it is guaranteed that $s\in \mathcal{PS}_N$. All such ST-cell $s$ form the partial pruned set $\mathcal{PPS}_N$, which can be used to create the artificial entity $e_v'$ with ST-cell set $\mathcal{S}_v'=\mathcal{S}_q-\mathcal{PPS}_N$. Evidently, $\mathcal{S}_v'$ may be slightly larger than $\mathcal{S}_v$, which leads to a larger upper bound $UB_N'$. However, as the hash value at the routing index is in general far larger than the other hash values in the group-level signature, $UB_N'$ is expected be very close to $UB_N$. In the experiment we utilize partial pruned sets to evaluate performance; details are given in Section \ref{experiments}.

As discussed in Theorem \ref{th:ancestor}, the pruned set of a descendant node contains that of its ancestor nodes. Therefore, in a specific branch of the MinSigTree, the upper bound can be gradually tightened before we reach the leaf nodes and check the contained entities.
\subsection{Search algorithm}
\label{sec:alg}
The search algorithm based on the early termination condition is given in Algorithm 2. We initialize the result as a priority queue sorted by association degree (from high to low), and start the search from the root of MinSigTree (Line 1) whose upper bound is set to $1$. We fetch the node with maximal $UB$ in the candidate list (Line 3) and insert all its child nodes into the candidate list (Line 8). Once we reach a leaf node, we calculate the exact association degree between the query node and all entities in this node, and update the result accordingly (Lines 10 - 13). The process terminates when either (1) we have identified $k$ entities, and the association degree between any of these $k$ entities and the query entity is no less than the maximal $UB$ of the remaining candidates (Lines 4 - 5), or (2) all leaves have been explored (Line 16). It is worth noting that the algorithm is applicable to all association degree measures as long as they satisfy the constraints of $d$ specified in Section \ref{prob-def}.
\begin{algorithm}[htb]
\caption{Top-$k$ query processing}
\label{alg:Framwork}
\begin{algorithmic}[1]
\Require
MinSigTree $T$, $k$, query entity $e$, measure $d$
\Ensure
$k$ most associated entities to $e$
\State \textbf{Initialization:} Result = \{\ \}, Candidate = \{root of $T$\};
\While {Candidate$\neq\emptyset$}
    \State $N = $ node with maximal $UB$ in Candidate;    
    \If {Result.minKey$\geq$N.UB and Result.size==$k$}
        \State return Result;
    \EndIf
    \If {$N$ is not leaf}
        \State Candidate = Candidate $\cup \{$all child nodes of $N\}$;
    \Else
        \State $\mathcal{E}_N = $ entities contained in $N$;
        \For {$e'$ : $\mathcal{E}_N$}
            \State $s = d(e,e')$;
            \State Result.update($e'$, $s$)
        \EndFor
    \EndIf
\EndWhile
\State \Return Result;
\end{algorithmic}
\end{algorithm}
\begin{exmp}
Let us again consider the MinSigTree in Figure \ref{fig:MinSigTree} as an example. We use a Dice similarity-based function as the measure of association degree: $d(e_i,e_j)=0.1\times \frac{\vert seq_i^1\cap seq_j^1\vert}{\vert seq_i^1\vert + \vert seq_j^1\vert}+0.9\times \frac{\vert seq_i^2\cap seq_j^2\vert}{\vert seq_i^2\vert + \vert seq_j^2\vert}$. Let $e_c$ be the query entity, and the Top-1 result is desired. As indicated in Example \ref{exmp:signature}, $seq_c^2=\{T_1L_3,T_2L_1\}$, $h_1(T_1L_3)=4$, $h_2(T_1L_3)=4$, $h_1(T_2L_1)=8$, $h_2(T_2L_1)=3$. We start the search from the root. For node $N_1$, as $3<h_1(T_1L_3)$ and $3<h_1(T_2L_1)$, we have $\mathcal{PPS}_{N_1}=\emptyset$, and thus the upper bound of $N_1$, $UB_{N_1}=1$; similarly $UB_{N_2}=1$. The candidate queue is $(1: \{N_1, N_2\})$. Since there is no remaining node at level 1, we dequeue $N_1$, the only child of which is $N_{12}$. As $7>h_2(T_1L_3)$ and $7>h_2(T_2L_1)$, we have $\mathcal{PPS}_{N_{12}}=\{T_1L_2,T_2L_1\}$, $UB_{N_{12}}=0.1\times 1 + 0.9\times 0=0.1$, where $1$ is the UB of the parent node of $N_{12}$, and $0$ corresponds to the fact that both query ST-cells are contained in $\mathcal{PPS}_{N_{12}}$. We then dequeue $N_{2}$. The first child of $N_2$ is $N_{21}$. As $4<h_1(T_1L_3)$ and $4<h_1(T_2L_1)$, $UB_{N_{21}}=1$. For node $N_{22}$, as $5>h_2(T_1L_3)$ and $5>h_2(T_2L_1)$, $UB_{N_{22}}=0.1\times 1 + 0.9\times 0=0.1$. The candidate queue becomes $(1:\{N_{21}\},0.1:\{N_{12},N_{22}\})$. We then dequeue $N_{21}$. Since $N_{21}$ is a leaf node, we calculate the actual association degree between $e_a$ and entities contained in $N_{21}$, and obtain $d(e_a,e_c)=0.5$. Since $d(e_a,e_c)> 0.1$, the algorithm returns $e_a$.
\end{exmp}

\section{Pruning Effectiveness Analysis}
\label{sec:him}
In this section, we introduce a hierarchical mobility model significantly extending a well-established single-level individual mobility (IM) model \cite{song2010modelling}. In addition, we theoretically analyze the pruning effectiveness of our algorithms using the proposed model.
\subsection{Individual mobility model}
\label{sec:onelevelim}
In the ensuing discussion, $\beta$, $\rho$, $\gamma$, $\alpha$, $\zeta$, $\mu$, and $\nu$ are all model parameters. 

For an entity $e$, the duration $\Delta t$ of each PI follows
\begin{equation}
P(\Delta t)\sim \vert\Delta t \vert^{-1-\beta},
\label{beta}
\end{equation}
which indicates  that the duration of each PI follows a power law distribution, i.e., entities tend to stay for a short duration at each base spatial unit than for a long period. 

When $e$ leaves the current base spatial unit, it will either take an exploratory jump to a new base spatial unit, or return to somewhere it has previously visited. The probability of taking an exploratory jump is
\begin{equation}
P_{new}=\rho S^{-\gamma},
\label{eq:newjump}
\end{equation}
where $S$ is the number of base spatial units visited. As $e$ visits more base spatial units, i.e., when $S$ increases, the probability of $e$ taking an exploratory jump decreases. 

The direction of an exploratory jump is selected randomly, and its displacement follows
\begin{equation}
P(\Delta r)\sim \vert\Delta r \vert^{-1-\alpha},
\label{eq:alpha}
\end{equation}
which stipulates that an entity tends to jump to some base spatial unit near its current position.

When taking a returning jump, the probability of returning to $l$ is proportional to the number of $e$'s previous visits to $l$. The visit frequency of $e$ to its $y$-{th} most visited base spatial unit follows
\begin{equation}
f_y\sim y^{-\zeta},
\label{eq:zeta}
\end{equation}
which indicates that most visits of an entity are to the few top-ranked base spatial units.

Given a duration $t$, the total number of distinct base spatial units visited by $e$ is
\begin{equation}
\label{im:s}
S(t)\sim t^\mu,
\end{equation}
and the mean squared displacement follows
\begin{equation}
\label{eq:msd}
\langle \Delta x^2(t)\rangle\sim t^\nu,
\end{equation}
which indicates that the longer the duration, the further $e$ will drift away from its starting position.

\subsection{Hierarchical individual mobility model}
\label{sec:hierarcahicalim}
The IM model in Section \ref{sec:onelevelim} describes human mobility patterns at the finest spatial level. However, AjPIs may occur at multiple levels. In this section, we give the general spatial units distribution patterns and aggregate the mobility pattern at the finest level into patterns at higher levels.

To ease analysis, we assume that the area of interest is a square with side length $L$, and that it is equally divided into a grid of non-overlapping cells where each cell is a square with side length $L_{bsu}$. Each base spatial unit corresponds to a cell in this grid. Therefore, there are $(\frac{L}{L_{bsu}})^2$ base spatial units in total. For the sp-index, the size of each spatial unit (i.e., the number of base spatial units contained therein) and the structure of the tree depend on two parameters:
\begin{itemize}
\item {\em Width}, i.e., the number of nodes at each level; and
\item {\em Relative density}, i.e., the relative sizes of nodes at the same level.
\end{itemize}
Intuitively, there are more spatial units at a finer level in the tree. Therefore, we assume that the width parameter follows a power law distribution w.r.t. level, i.e.,
\begin{equation}
\label{eq:nl}
    W_l= Q\cdot l^a,
\end{equation}
where $l\in [1,m]$ is the level, $a$ is a tunable parameter, and $Q=(\frac{L}{L_{bsu}})^2/{m^a}$ serves as a normalization factor.

In most cases the nodes at the same level have varying sizes, e.g. business districts usually have more buildings than rural areas. Therefore, we use the following power law distribution to model the relative sizes of nodes at level $l$:
\begin{equation}
    \label{eq:ni}
    D^i_l= W_l\cdot R\cdot i^b,
\end{equation}
where $i\in [1,W_l]$ is the index of nodes at level $l$, $b$ is a tunable parameter, and $R=1/{\sum_{i=1}^{W_l}i^b}$ is a normalization factor.

With parameters $L$, $L_{bsu}$, $a$ and $b$, we can obtain the number of spatial units and also the size of each spatial unit at any level. Next we demonstrate how distributions introduced in Section \ref{sec:onelevelim} modeling mobility at the finest level can be extended and supplemented with other necessary distributions to derive a hierarchical mobility model.


Let $U$ be a spatial unit at level $l$ which contains a set of base spatial units $\mathcal{S}_U$. An exploratory jump of an entity takes place when (1) the entity jumps to a new base spatial unit; and (2) the new base spatial unit is contained in a spatial unit previously not visited, at level $l$. The probability of the first condition is given in Equation~(\ref{eq:newjump}); the probability of the second condition, referred to as $P_{out}$, can be computed by
\begin{equation}
\label{eq:mlnewjump}
    P_{out}(U) = \frac{n_{visited}^U}{n_{reachable}^U}\sum_{s\in \mathcal{S}_U}\frac{1}{\vert \mathcal{S}_U\vert}H(s),
\end{equation}
where $n_{reachable}^U$ denotes the number of spatial units within one jump's distance from $U$, $n_{visited}^U$ denotes the number of spatial units visited among these reachable ones, $s$ is a base spatial unit in $\mathcal{S}_U$, and $H(s)$ denotes the probability of jumping outside $U$ from $s$. It is evident that $H(s)$ is a function of the distance from $s$ to the boundary of $U$ as well as the jump distance distribution given in Equation~(\ref{eq:alpha}). Therefore, the probability of taking an exploratory jump to a new spatial unit, $P_{new}'$, is
\begin{equation}
    P_{new}'(U) = P_{new}\times P_{out}(U)
\end{equation}


Since spatial units at higher levels have varying sizes and ranges, it is essential to derive the probability of an entity having visited unit $U$ (the size of which is $\vert \mathcal{S}_U\vert$) within time $t$, $P_U(t)$.
\begin{equation}
\label{eq:mlst}
    P_U(t)=\frac{\vert \mathcal{S}_U\vert}{\vert \mathcal{S}\vert}+ \sum_{U'}M(U,U',t),
\end{equation}
where $\mathcal{S}$ denotes the set of base spatial units, $U'$ denotes some other spatial unit at level $l$. To derive this probability we consider two cases: the starting position of the entity is within $U$ or it is not. The probability of the former case is $\frac{\vert \mathcal{S}_U\vert}{\vert \mathcal{S}\vert}$. For the latter case, the starting position can be within any other spatial unit, $U'$. $M(U,U',t)$ describes the probability of an entity starting from unit $U'$ having visited $U$ after time $t$, which can be inferred by the mean square displacement distribution given in Equation~(\ref{eq:msd}).

The visit frequency of an entity to its $y$-th most visited base spatial unit is given in Equation~(\ref{eq:zeta}). At higher levels, the visit frequency rank, $y$, reflects not only personal preference, but also unit characteristics: spatial units containing more base spatial units are likely to be top-ranked. Therefore, we can safely assume that the visit frequency follows the same distribution at higher level, where $y$ now describes the rank of the visit frequency to a particular spatial unit.  
\subsection{Analysis of pruning effectiveness}
\label{sec:analysis}
The model proposed in Section \ref{sec:hierarcahicalim} enables us to simulate the movements of entities, estimate the overlap between the digital traces of any entities at all levels, with which we can calculate the expected association degree between an entity and its $k$ most associated entities, $d_e$. Thus we can discard all branches whose upper bound is smaller than $d_e$. With more branches discarded, answering the query will be more efficient. Here we formally define pruning effectiveness (PE):
\begin{definition}[Pruning Effectiveness]
Given a set of entities $\mathcal{E}$, a query entity $e$, an association degree measure $d$, and a searching strategy $S$, if $S$ accurately answers a top-$k$ query w.r.t $\mathcal{E}$, $e$, and $d$ by checking entities in set $\mathcal{E}'$ ($\mathcal{E}'\in \mathcal{E}$) only, then the pruning effectiveness of $S$ is $\frac{ \vert \mathcal{E}'\vert - k}{\vert \mathcal{E}\vert}$. 
\end{definition}

The average PE of is obtained averaging the PE of the top-$k$ query answers over multiple entities.

Evidently, the UB of a child node on the MinSigTree cannot be larger than that of its parent nodes. Therefore, we can estimate PE by computing the percentage of leaf nodes on MinSigTree whose UBs are larger than $d_e$.

Suppose that the total number of base spatial units is $n$ and the duration is $t$, the range of hash functions is thus $[0, n\times t-1]$. For entity $e_a$ with ST-cell set sequence $seq_a$ and signatures $sig_a$, the probability of $sig_a^m[u]=i$ is
\begin{equation}
  p(sig_a^m[u]=i)=\sum\limits_{x=1}^{\vert seq_a^m\vert}C_{\vert seq_a^m\vert}^x(\frac{1}{n\times t})^x(\frac{n\times t-i}{n\times t})^{\vert seq_a^m\vert-x}
\end{equation}
The condition of $sig_a^m[u]=i$ is that, $\exists \mathcal{S}_a \subset seq_a^m$, $\mathcal{S}_a \neq \emptyset$, s.t. $\forall s\in \mathcal{S}_a$, $h_u(s) = i$, and $\forall s'\in seq_a^m-\mathcal{S}_a$, $h_u(s') > i$. We assume that $\vert \mathcal{S}_a\vert = x$, the probability of which is $C_{\vert seq_a^m\vert}^x(\frac{1}{n\times t})^x$, then all remaining ST-cells take hash values larger than $i$, the probability of which is $(\frac{n\times t-i}{n\times t})^{\vert seq_a^m\vert-x}$. By grouping entities with the MinSigTree, the signature of a node $N$, $SIG_N$, satisfies $p(SIG_N[u]=i)\approx p(sig_a^m[u]=i)$ (equal when $N$ only contains $e_a$).

Let $r$ be the routing index of $N$, then the probability of $SIG_N[r]=i$ is
\begin{gather}
\label{eq:ri}
    p(SIG_N[r]=i)  = \sum_{x=1}^{n_h}C_{n_h}^xp(SIG_N[u]=i)^x p(SIG_N[u]<i)^{n_h-x},\nonumber \\
    p(SIG_N[u]<i)  = \sum_{x = 0}^{i-1} p(SIG_N[u]=x)
\end{gather}

With the knowledge of $p(SIG_N[r]=i)$ we can estimate the value distribution of all leaves. Assume that range $[0,n\times t - 1]$ is divided into $n_r$ consecutive equal-sized sub-ranges $R$, then we use $V[j]$ to denote the percentage of leaves whose value on the routing index is bounded by $R[j]$, $0\leq j< n_r$.

Let $n_c$ be the minimal number of ST-cells shared by entities with association degree larger than $d_e$. For node $N$, if $\exists \mathcal{S}_{ap}\in seq_a^m $, $\vert \mathcal{S}_{ap}\vert \geq n_c$, s.t. $\forall s\in \mathcal{S}_{ap}$, $s\notin \mathcal{PS}_N$, then $N$ cannot be discarded.

Since hash functions are selected randomly, the hash values of all ST-cells are independent. Suppose that $SIG_N[r]$ is bounded by $R[j]$, then the probability that $N$ cannot be discarded is
\begin{equation}
\label{eq:pcheck}
    q(R[j])=\sum_{x=n_c}^{\vert seq_a^m\vert}C_{\vert seq_a^m\vert}^{x}(\frac{n\times t - 1 - R[j]}{n\times t - 1})^x(\frac{R[j]}{n\times t - 1})^{\vert seq_a^m\vert - x}
\end{equation}
PE can thus be calculated with the following equation:
\begin{equation}
    PE = \sum_{j=0}^{n_r}V[j]q(R[j])
\end{equation}
\section{Experiments}
\label{experiments}
In this section, we present a thorough experimental evaluation of our approach utilizing synthetic and real datasets, varying parameters of interest to explore the sensitivity of our proposal as well as PE trends.
\subsection{Settings}
\label{settings}
\textbf{Environment}. The experiments are conducted on an Amazon Web Service EC2 instance, with a 30 core 2.3GHz Xeon CPU, 120GB of RAM, and ITB EBS Throughput Optimized HDD (maximal throughput 1,750MiB/s). The programming language is Java (version 1.8.1).

\textbf{Datasets}. We employ both real and synthetic datasets in our evaluation. Synthetic data are used as it is easy to vary parameters for sensitivity analysis. The synthetic dataset (referred to as SYN in the sequel) is generated by the hierarchical IM model in Section \ref{sec:him} with varying values of the parameters $\alpha$, $\beta$, $\gamma$, $\zeta$, $\rho$, $a$, $b$ and $m$. Unless otherwise specified,we set $\alpha=0.6$, $\beta=0.8$, $\gamma=0.2$, $\zeta=1.2$, $\rho=0.6$, which correspond to the normal mobility pattern (as per \cite{song2010modelling}), and $a=2$, $b=2$, $m=4$ ($a$ and $b$ usually take values in the range $[1,2]$ in real datasets\footnote{https://data.cityofnewyork.us/City-Government/Points-Of-Interest/rxuy-2muj}, and $4$ is the typical hierarchical level in a city). The sensitivity to these parameters governing data characteristics is evaluated in Section \ref{sec:im}. The locations in the data are drawn from a set of 9 equal-sized sp-indexes with $250$K locations in total. The data consists of the digital traces of 100M entities for a period of 30 days. The real dataset (referred to as REAL) is a WiFi hotspot handshaking data set provided to us by a large telecommunications provider and includes 30 million mobile devices and 76,739 WiFi hotspots. The hotspots are organized into a 4-level sp-index.

The data distribution is depicted in Appendix \ref{append:datades}.

\textbf{Association degree measure}. There are two properties any association degree measure (ADM in sequel) must possess (as discussed in Section~\ref{prob-def}), namely monotonicity with respect to AjPI level and duration. For purposes of exposition we utilize the following extensible function as the ADM:
\begin{equation}
\label{eq:measure}
    d(e_a,e_b)=\frac{\sum\limits_{l=1}^{m}l^{u}
    (\frac{\vert\mathcal{P}_{ab}^l\vert}{\vert \mathcal{P}^l_a\vert+\vert \mathcal{P}^l_b\vert})^{v}
    }{max},
\end{equation}
where $max$ is a normalization factor guaranteeing the score falls into the range $[0,1]$, $\vert \mathcal{P}^l_{ab}\vert$ denotes the total duration of all level $l$ AjPIs in set $\mathcal{P}_{ab}$, and $u>0$ and $v>0$ are parameters that can be tuned.

Note that measures such as Jaccard, Dice and Cosine Similarity can be readily applied as well as they all share the two properties required by our association degree measure. We compare the ranking results of the proposed ADM to those of several widely-adopted set similarity measures in Appendix \ref{appen:comp}. The results indicate that the proposed ADM compares favourably in terms of its ranking results to Jaccard, Dice and Cosine Similarity, when $v\in[0.5,2]$ in Equation (\ref{eq:measure}), and thus we utilize values in this range during experiments.

\subsection{Baseline approach}
\label{sec:baseline}
We consider the following approach based on locality as the baseline approach for comparison purposes. At each level, we treat an ST-cell set of an entity as a transaction, each ST-cell as an item, and utilize frequent pattern mining techniques to find those frequently co-occurring ST-cells. As a result, ST-cells are partitioned into clusters, where each cluster is expected to contain ST-cells that are close to one another temporally and spatially. If there are $n$ clusters in total, then we can assign each entity an $n$-bit vector, where the $i$-th bit in the vector of $e$ equals 1 if $e$ has presence in at least one ST-cell contained in cluster $i$, and 0 otherwise. We can thus use a bit-map to organize all entities. Given a query entity $e_q$, we compute an ADM upper bound between $e_q$ and all bit-vectors. We start searching from the entities indexed by the vector with the highest UB, and continue until $k$ entities are found where the minimal ADM of the $k$ entities is already greater than the UBs between $e_q$ and all remaining vectors. 

The major drawback of such an approach is that in practice ST-cells show low degrees of locality, e.g., people living in the same neighborhood may work in different companies spread across the city, which makes it very difficult to identify frequently co-occurring ST-cells. The direct consequence is that the clusters obtained demonstrate strong coupling and the bit vectors cannot capture the PI patterns of entities well. Therefore, the upper bound is loose as will be discussed later in Section \ref{sec:k}.

\subsection{Sensitivity to the number of hash functions}
\begin{figure}
\centering
\subfloat[REAL data]{\includegraphics[width=.45\columnwidth, height = 2cm]{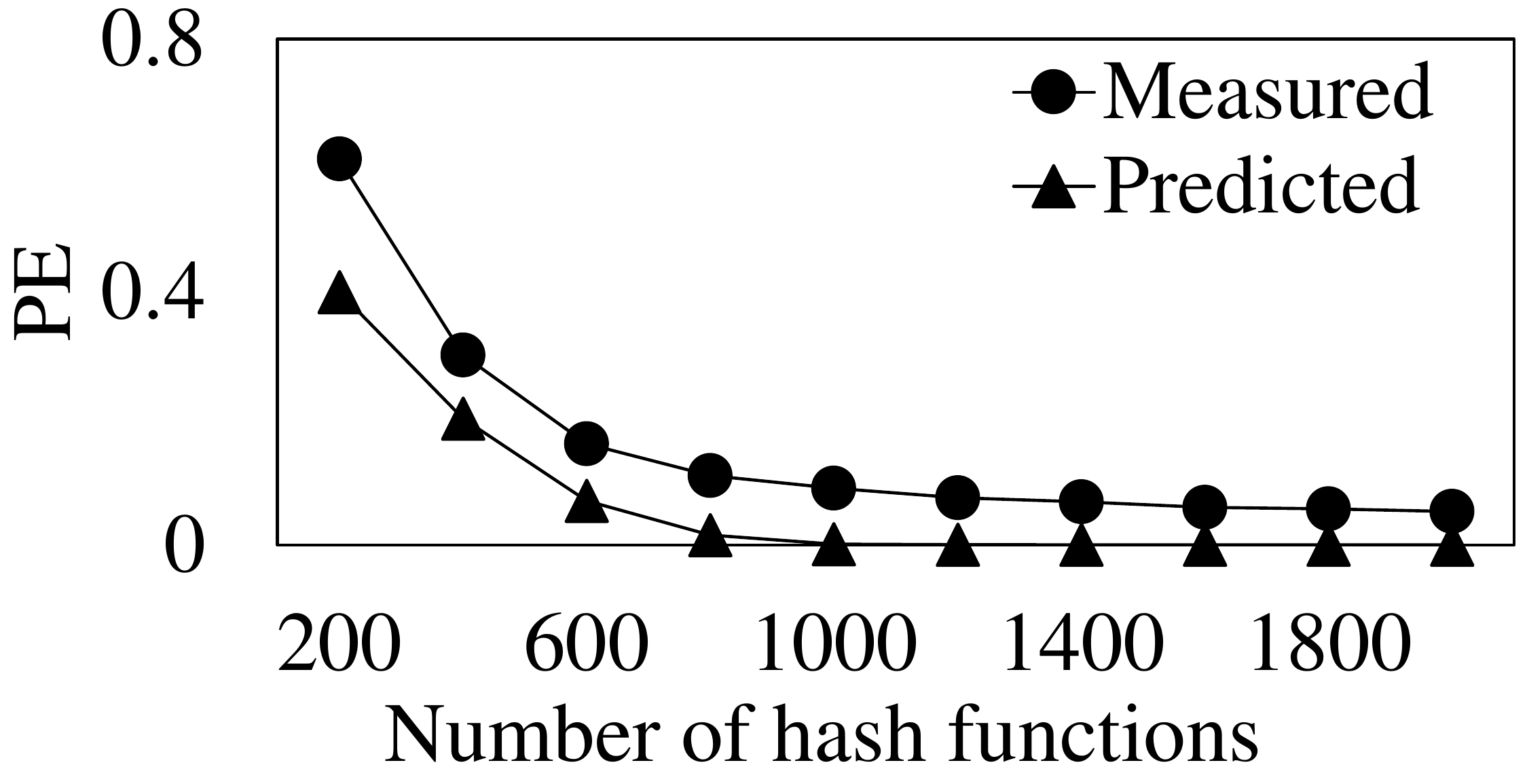}}\quad
   \subfloat[SYN data]{\includegraphics[width=.45\columnwidth, height = 2cm]{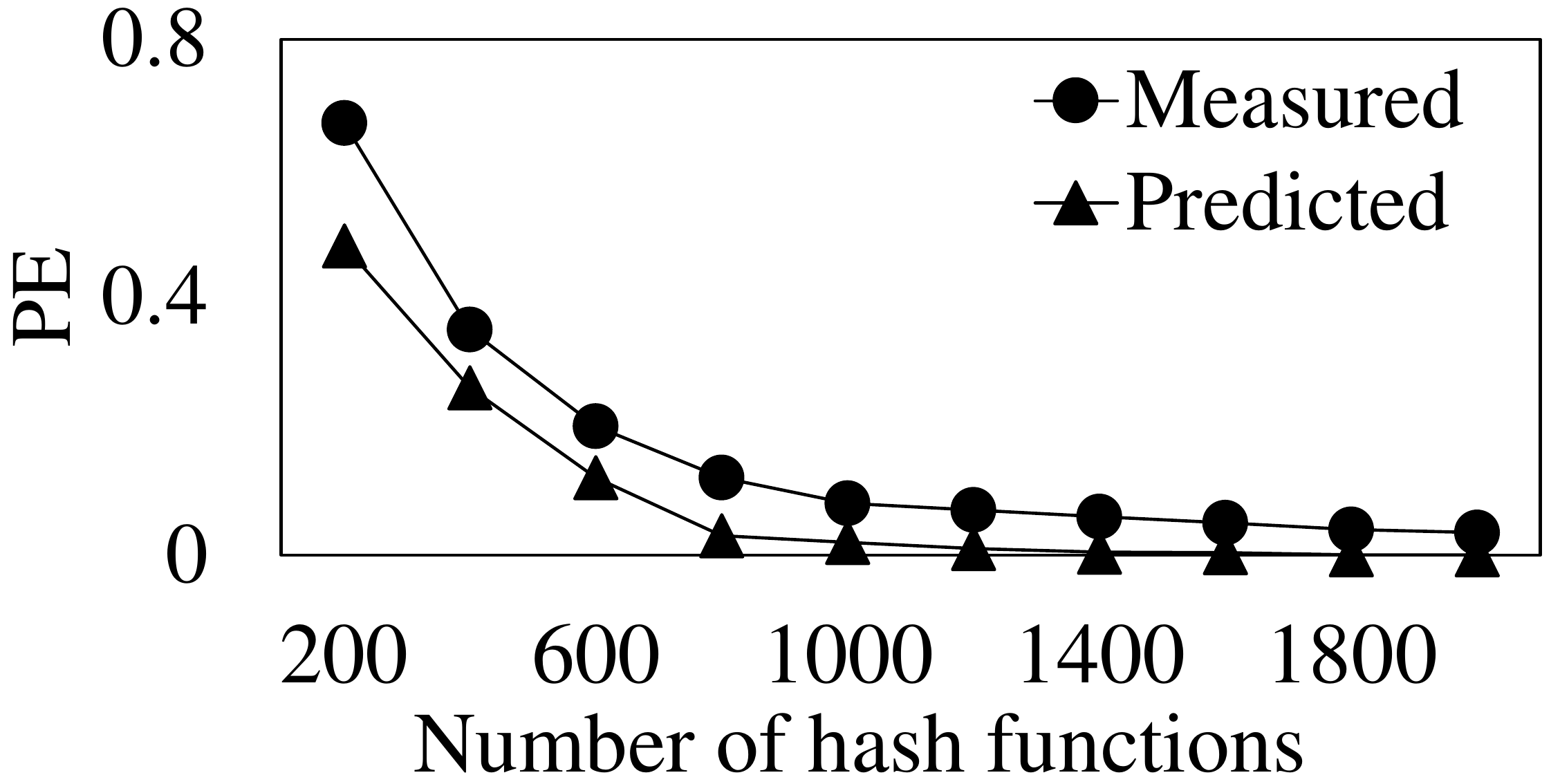}}
   \vspace{-4mm}
\caption{PE vs. the number of hash functions}
\label{pruning}
\end{figure}
\label{sec:parameter}
PE is closely related to the number of hash functions utilized to compute the signatures, $n_h$. We thus evaluate the PE of the proposed approach by varying $n_h$, and compare the measured PE in the experiment with the one predicted for the model of Section \ref{sec:analysis}. The results are presented in Figure \ref{pruning}.

From the result one can observe that the MinSigTree provides high PE with more hash functions. The reason is that, compressing the large number of ST-cells into a low-dimensional space makes entities less unique, or even indistinguishable. With more hash functions employed, signatures can better summarize the PIs of entities and thus only closely associated entities will be placed in the same group. Diminishing returns occur when the number of hash functions reaches 1,000, as each entity has become unique enough that further employment of hash functions does not change the grouping.

As Figure \ref{pruning} shows, the predicted PE is slightly better than measured, primarily for the following reasons:
\begin{itemize}
\item Spatial units in the hierarchical IM model are assumed to be rectangles for analysis purposes, while in practice units can be in any shapes. As a result, the mobility patterns at higher levels diverge from the model;
\item It is assumed that the hash values are uniformly distributed on the range, which is not always the case in practice.
\end{itemize}
\subsection{Sensitivity to data characteristics}
\label{sec:im}
\begin{figure}
   \centering
   \subfloat[][]{\includegraphics[width=.45\columnwidth, height = 2cm]{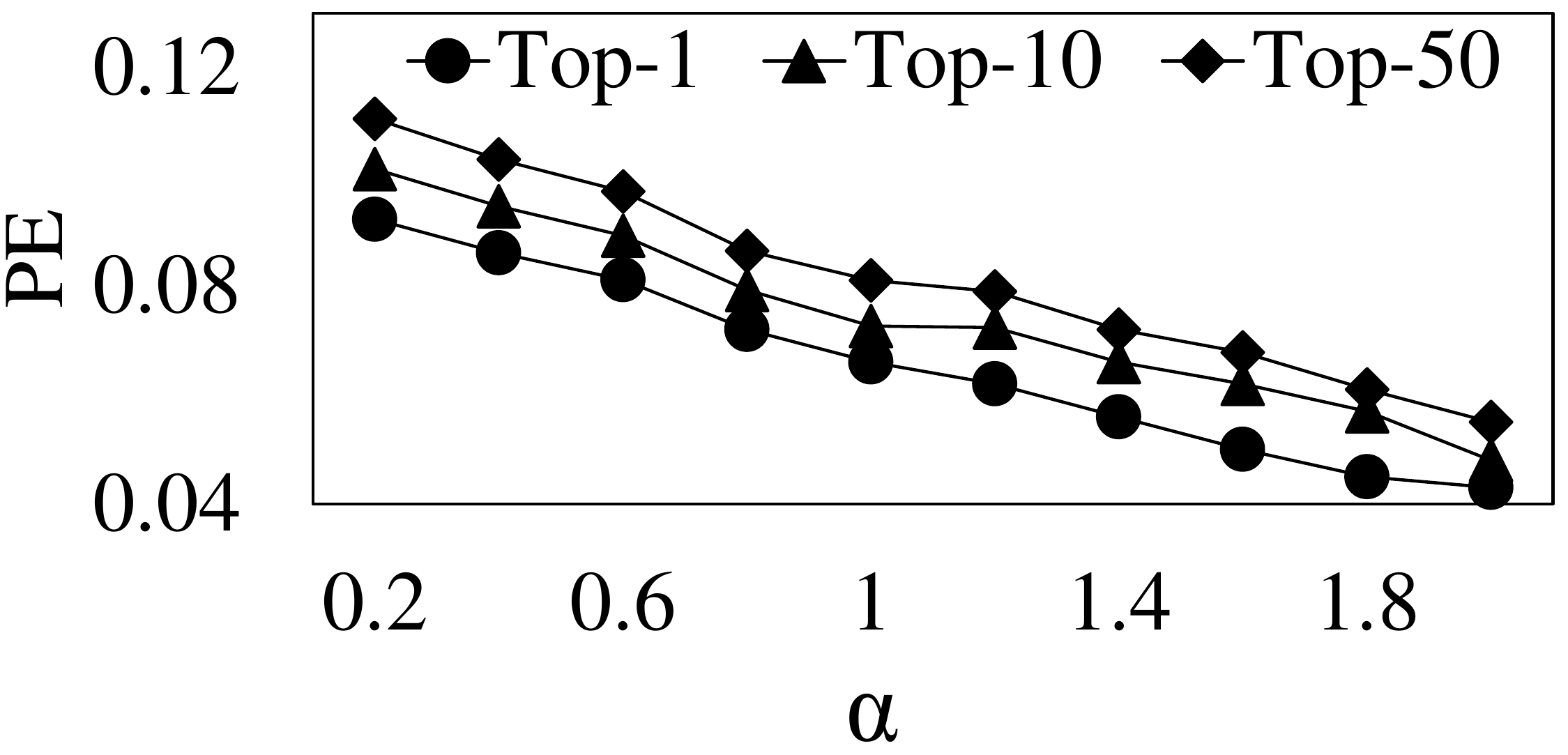}}\quad
   \subfloat[][]{\includegraphics[width=.45\columnwidth, height = 2cm]{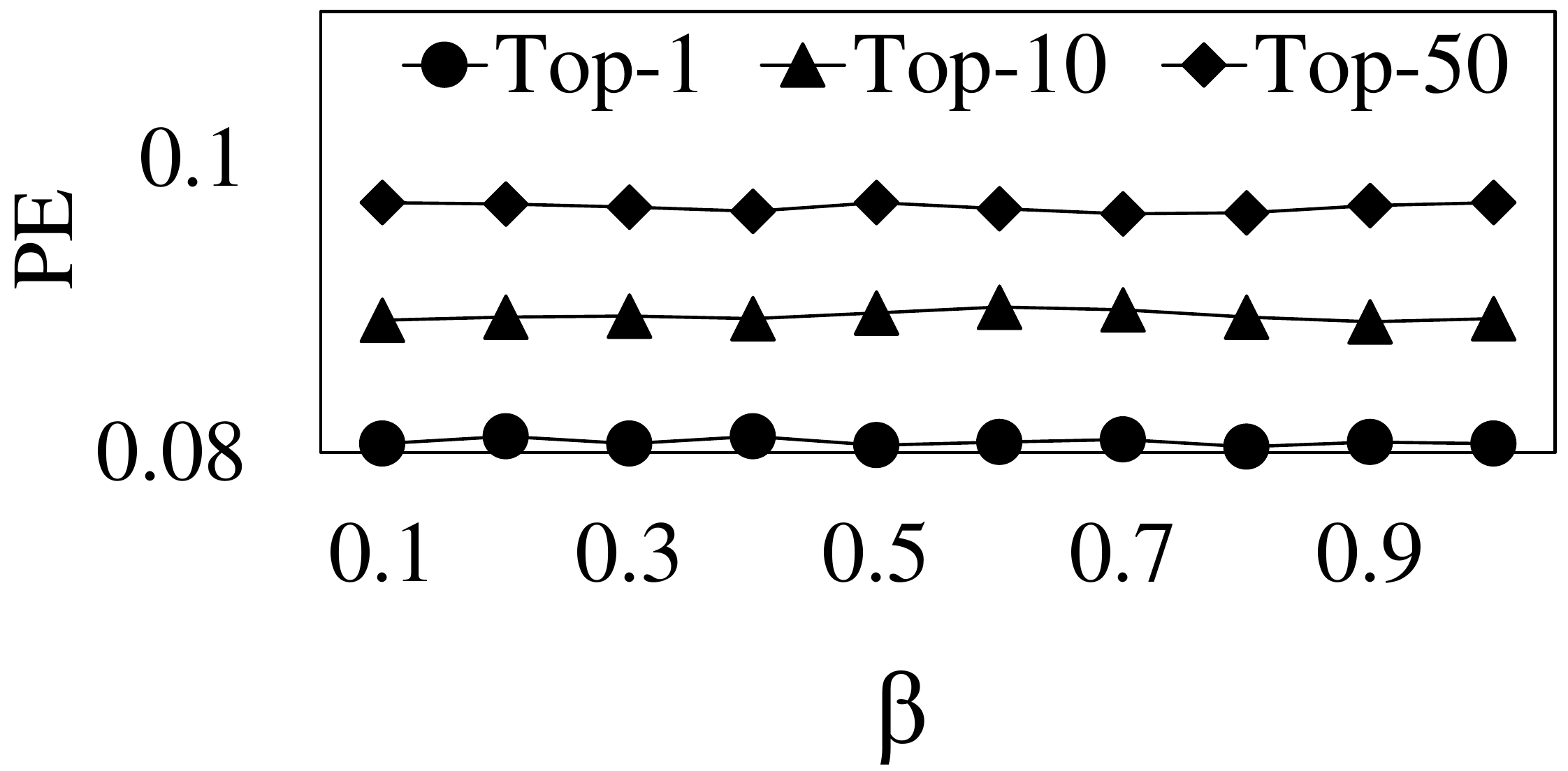}}\\
   \subfloat[][]{\includegraphics[width=.45\columnwidth, height = 2cm]{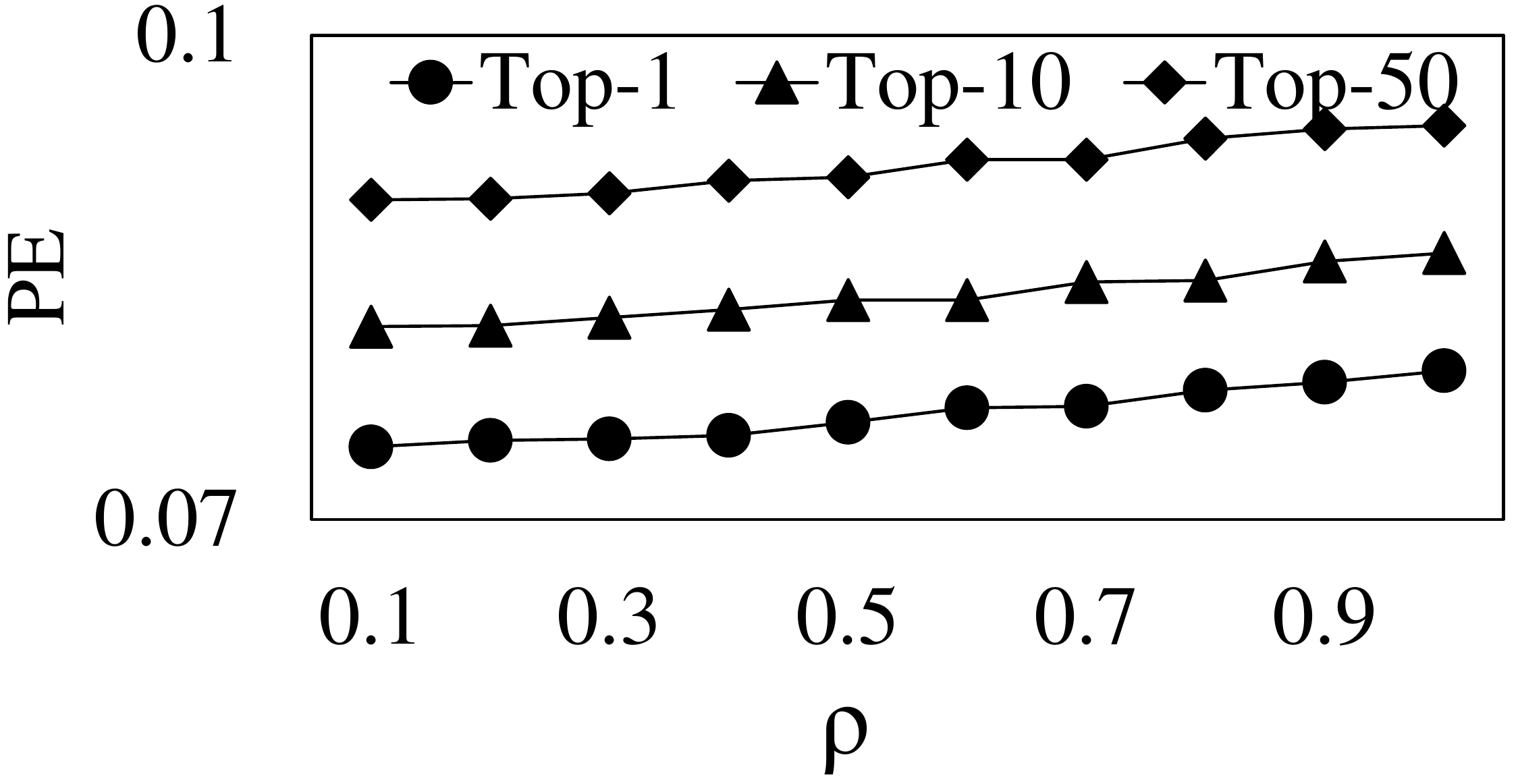}}\quad
   \subfloat[][]{\includegraphics[width=.45\columnwidth, height = 2cm]{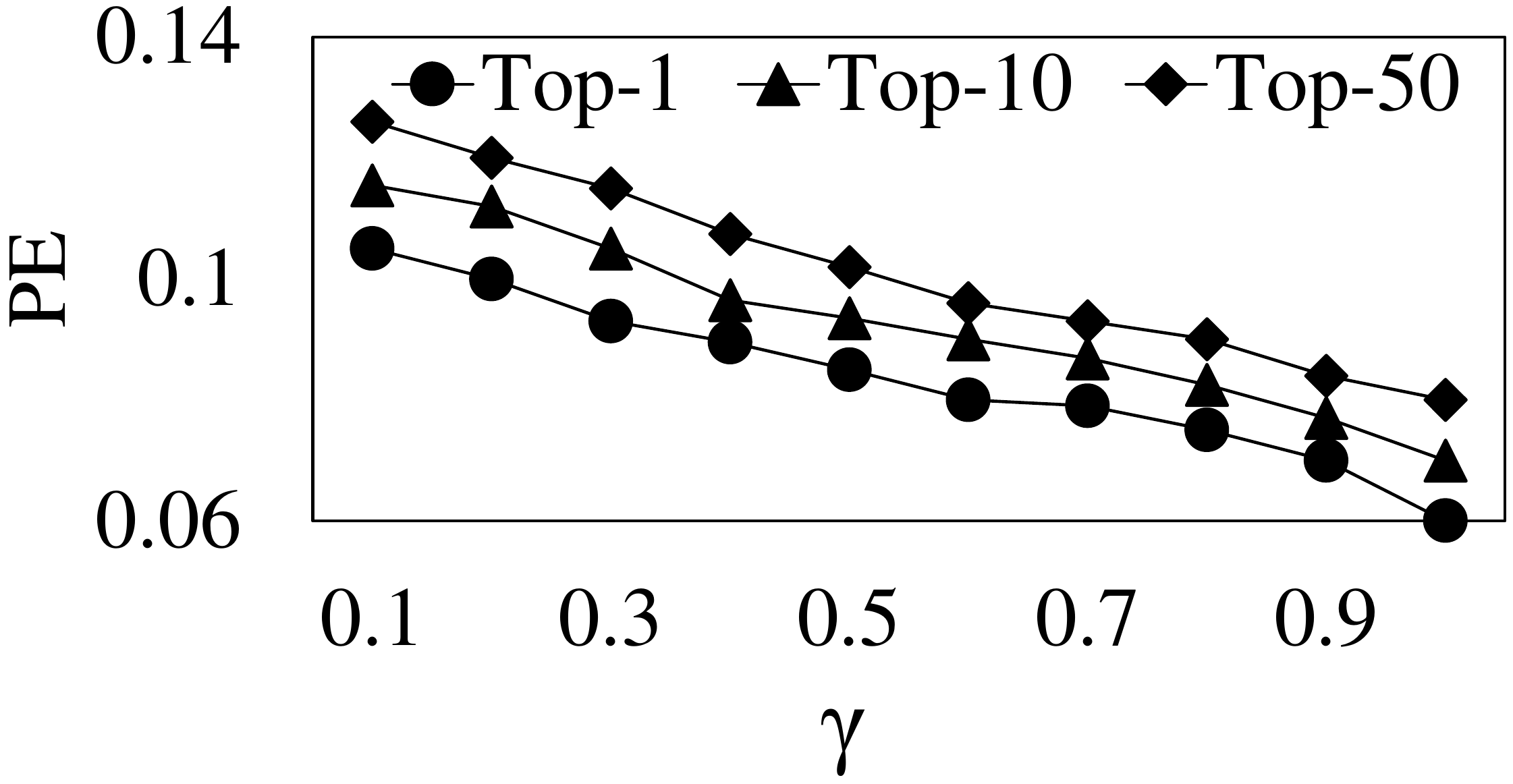}}\\
   \subfloat[][]{\includegraphics[width=.45\columnwidth, height = 2cm]{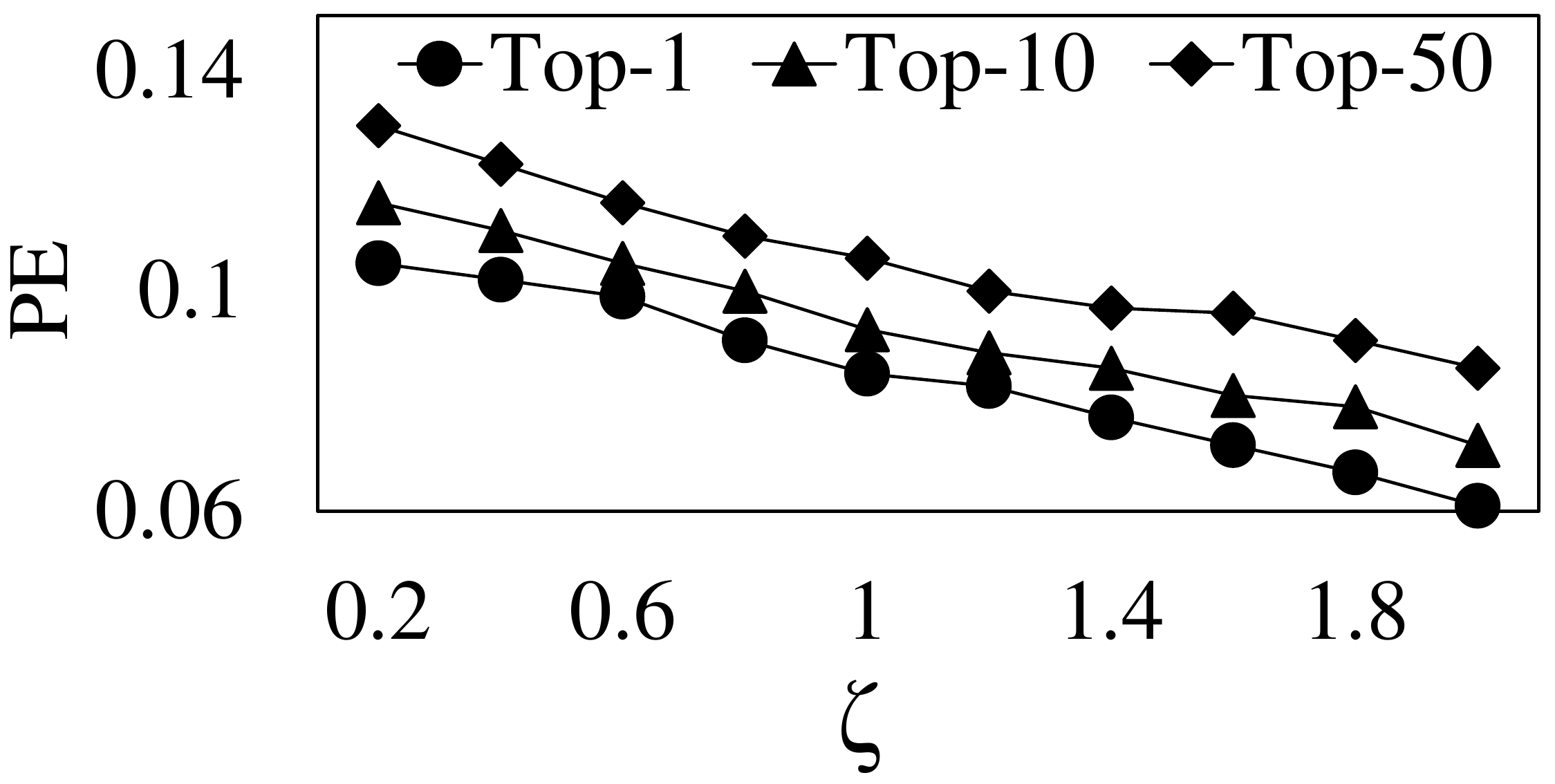}}\quad
   \subfloat[][]{\includegraphics[width=.45\columnwidth, height = 2cm]{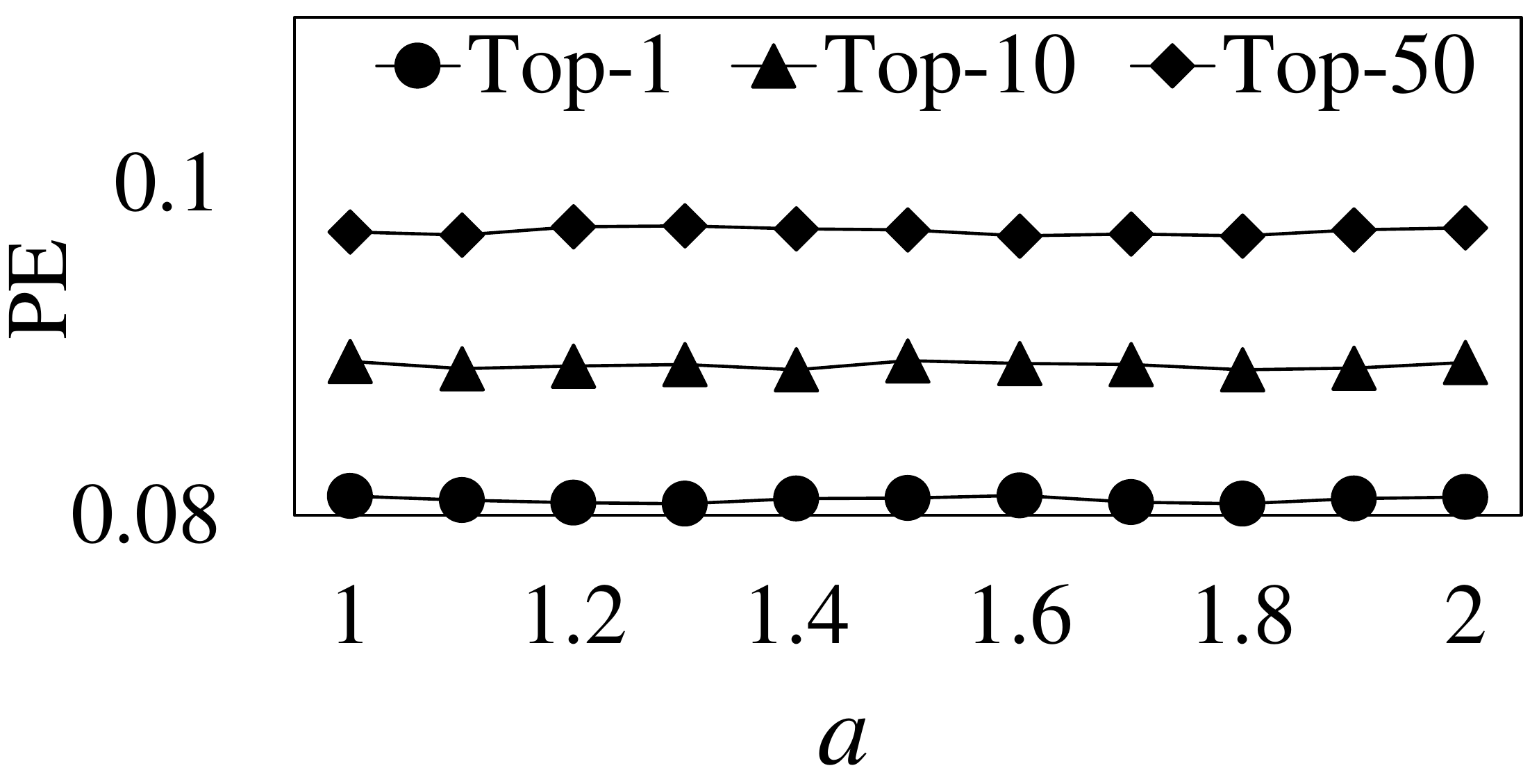}}\\
   \subfloat[][]{\includegraphics[width=.45\columnwidth, height = 2cm]{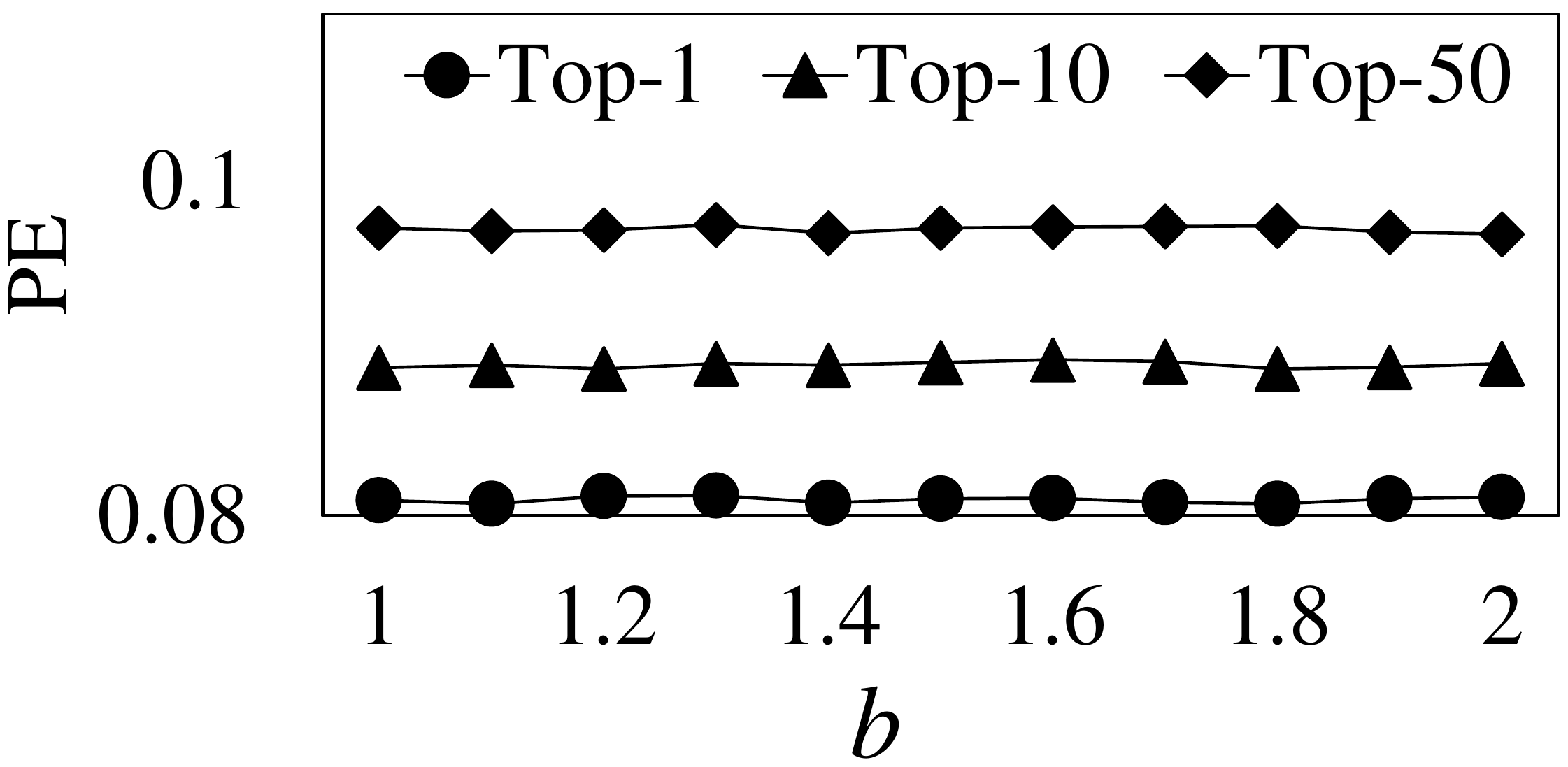}}\quad
   \subfloat[][]{\includegraphics[width=.45\columnwidth, height = 2cm]{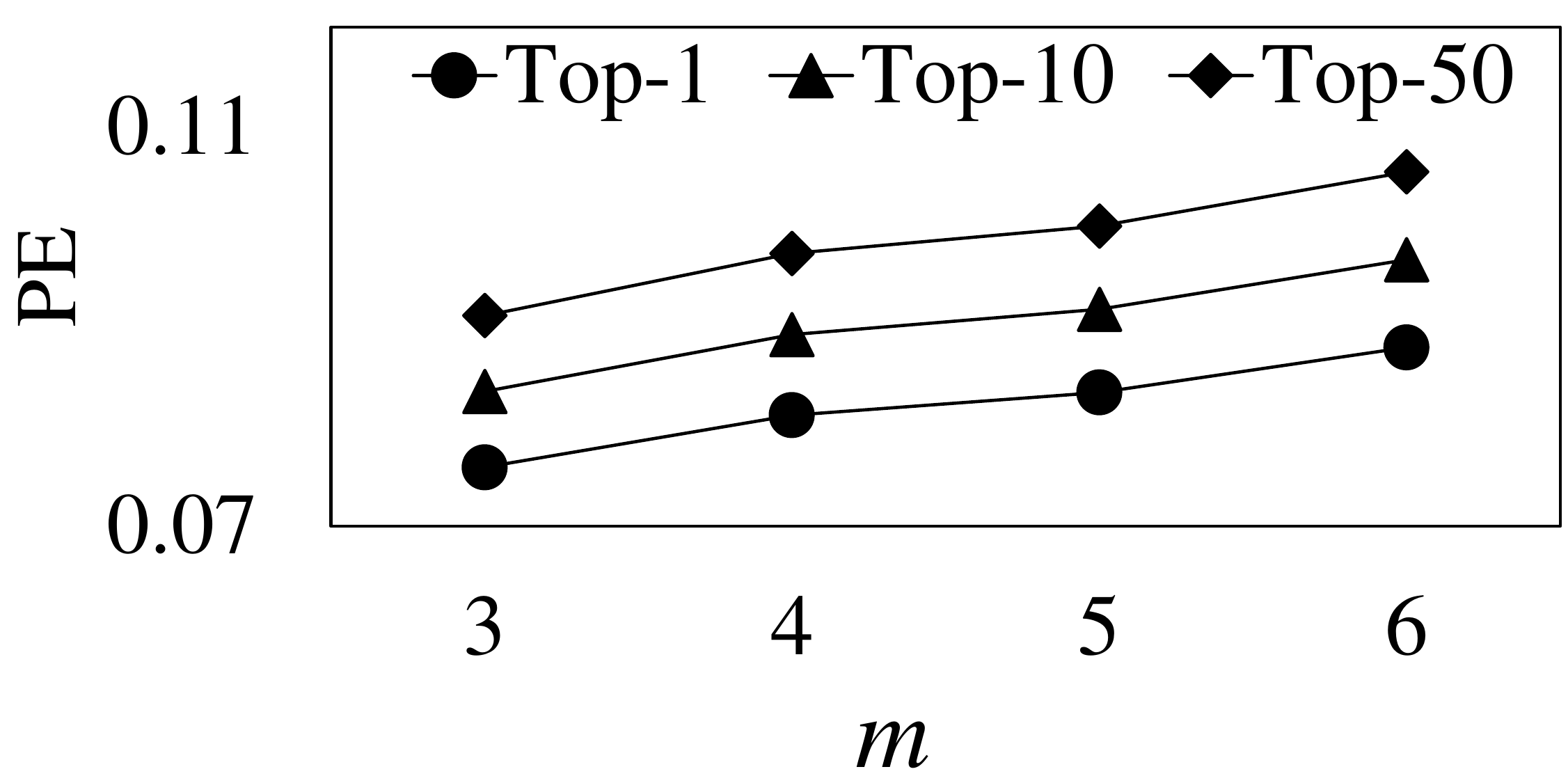}}
   \vspace{-4mm}
   \caption{PE vs. data characteristics}
   \label{im}
\end{figure}

We evaluate the PE under different mobility patterns and location distributions by varying all parameters in the hierarchical IM model. The hierarchical IM model involves a large number of parameters, each controlling different aspects of human mobility or location distribution. As such we vary one parameter each time and fix other parameters to the value associated with normal patterns (as per \cite{song2010modelling}) to investigate the individual influence of different parameters on performance. The results of answering Top-1, Top-10, and Top-50 queries with 2,000 hash functions under different data characteristics are presented in Figure \ref{im}.

One can observe that curves in Figure \ref{im}(a) show a descending trend, as $\alpha$ controls the movement locality in the following way: as $\alpha$ increases, an entity is more likely to jump to locations in proximity when it leaves the current position. A higher level of locality will produce more closely associated entities, and thus lead to better performance.

Curves in Figure \ref{im}(b) demonstrate little variation, which indicates that the approach is not sensitive to the expected duration of each presence instance. This is because we partition PI into ST-cells, and consider the digital traces of an entity as a set of ST-cells. As a result, whether these ST-cells are consecutive in time or not has no influence on PE. 

Parameters $\rho$ and $\gamma$ together control the tendency of an entity to return to some previously visited location. With smaller $\rho$ and larger $\gamma$, entities visit fewer locations in total, which increases the locality. Therefore, Figure \ref{im}(c) depicts an ascending trend and Figure \ref{im}(d) a descending trend. $\rho$ acts as a linear parameter, while $\gamma$ is on the exponent; therefore curves in Figure \ref{im}(d) appear steeper than in Figure \ref{im}(c).

Similarly, Figure \ref{im}(e) demonstrates a descending trend, as $\zeta$ influences the locality by controlling the visit frequency distribution of an entity to locations. With higher $\zeta$, most visits are to a few most frequently visited locations, while with lower $\zeta$, visits are more uniformly distributed. 

Curves in Figure \ref{im}(f) and (g) depict little variation, indicating that good PE can be achieved under any spatial distribution patterns. As is clear from the search algorithm, we touch the records of entity $e$ only if the PI patterns of $e$ resembles the PI patterns of the query entity at all sp-index levels. Although the values of $a$ and $b$ influence spatial units distribution at higher levels, base spatial unit numbers and distributions in the explored area are always constant, which means that the PI patterns of entities at the finest level do not change. As a result, groupings at level $m$ of the MinSigTree remain unchanged under different values of $a$ and $b$.

From Figure \ref{im}(h) we observe that the approach performs better with smaller $m$, i.e., fewer levels in the hierarchy. The reason is that  with more spatial levels, more entities form AjPIs with each other, and thus the search space grows. As an example, if we assume that the spatial hierarchy is city-street-district-building, then if $m=1$, we only consider AjPIs at the building level, while with $m=2$ we consider AjPIs at both the building level and the street level, etc.
\subsection{Sensitivity to ADM parameters}
\label{sec:measure}
\begin{figure}
   \centering
   \subfloat[REAL data]{\includegraphics[width=.45\columnwidth, height = 2cm]{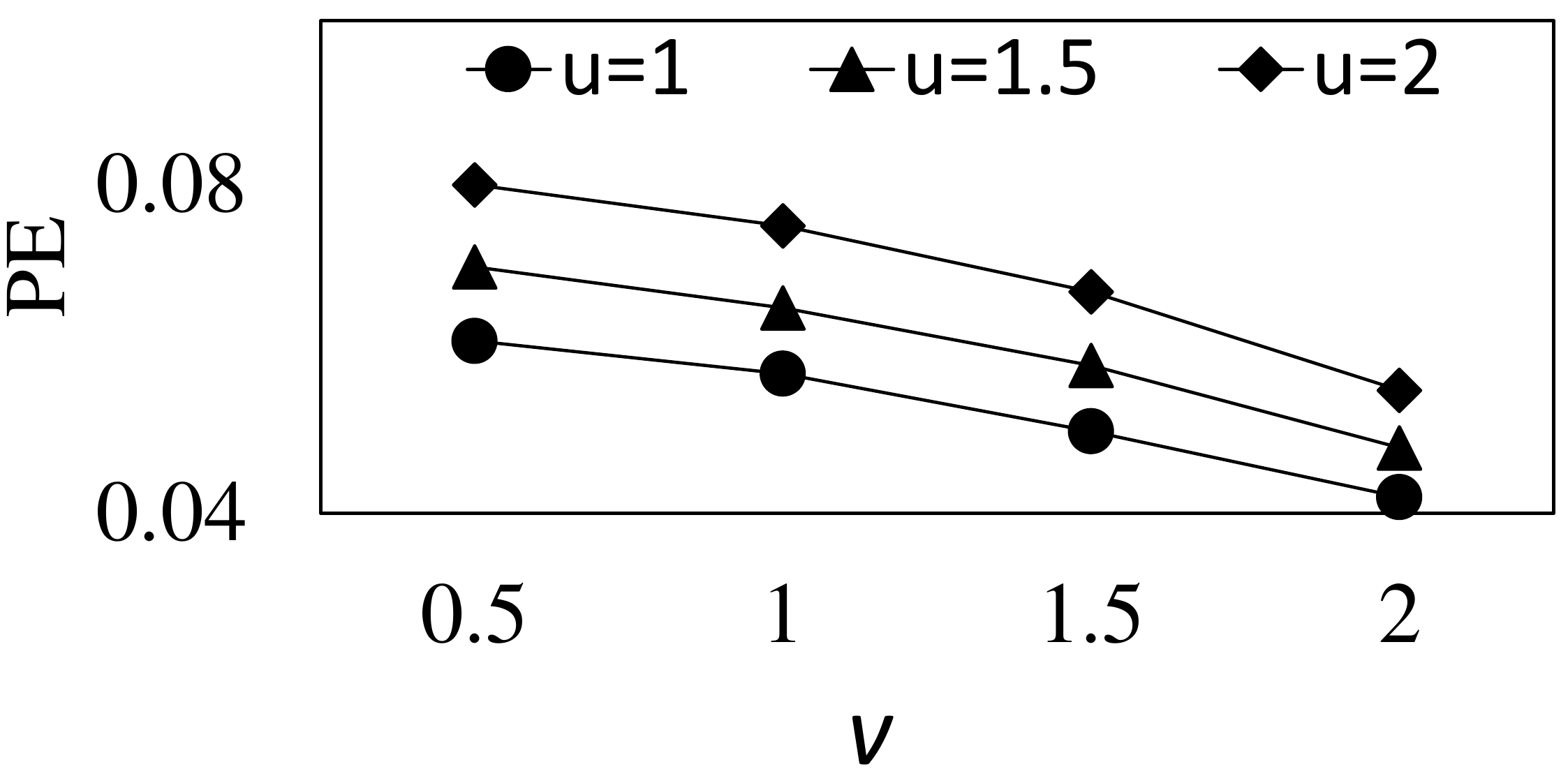}}\quad
   \subfloat[SYN data]{\includegraphics[width=.45\columnwidth, height = 2cm]{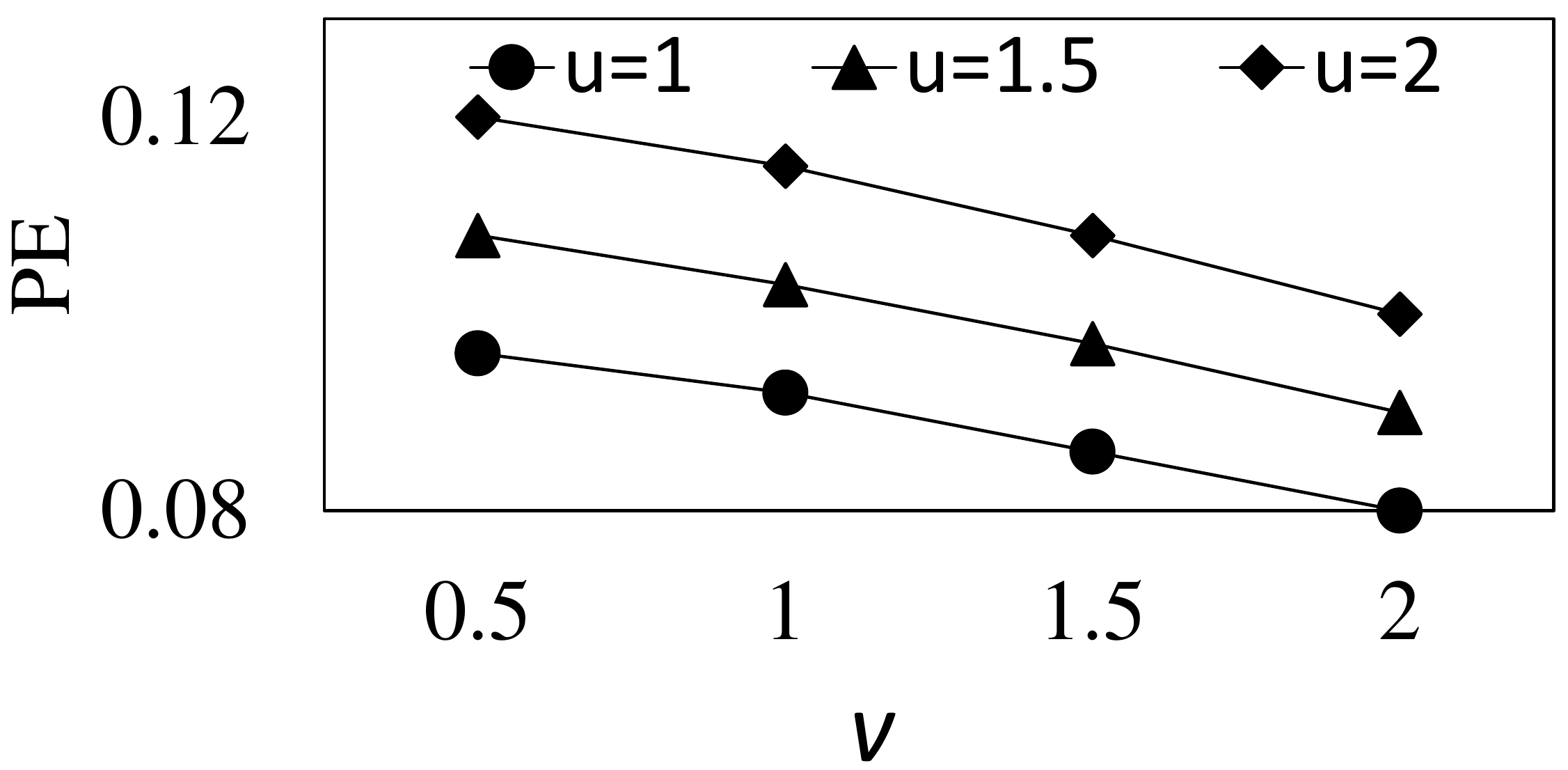}}
   \vspace{-4mm}
   \caption{PE vs. ADM parameters}
   \label{fig:measure}
\end{figure}

Values of $u$ and $v$ defined in the ADM of Section \ref{settings} provide different weights to AjPI level and duration when selecting associated entities. The PE under different ADM parameter values are presented in Figure \ref{fig:measure}.

As is clear from Figure \ref{fig:measure}, smaller $u$ (level parameter) and larger $v$ (duration parameter) yield high PE in both data sets. The reason is that, while ST-cells contain timestamps, they do not contain level information. Since signatures are computed based on ST-cells, the AjPI level is not encoded into the signature. As a result, entities sharing AjPIs for longer duration are more likely to have similar signatures than entities sharing AjPIs at finer levels. The results reveal that the approach performs better in cases where duration is the dominant factor of the association degree between entities.

\subsection{Sensitivity to memory size}
\begin{figure}
   \centering
   \subfloat[REAL data]{\includegraphics[width=.45\columnwidth, height = 2cm]{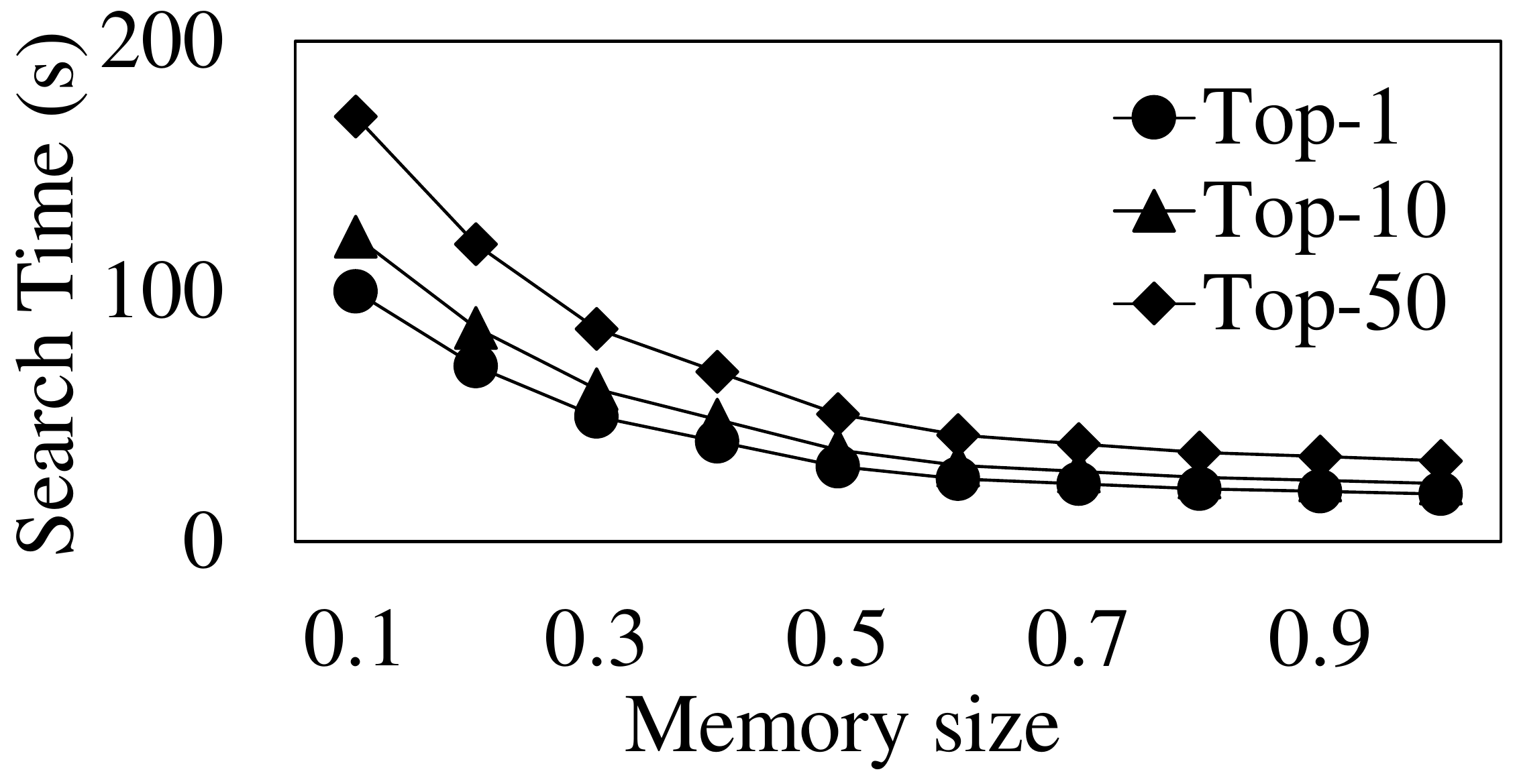}}\quad
   \subfloat[SYN data]{\includegraphics[width=.45\columnwidth, height = 2cm]{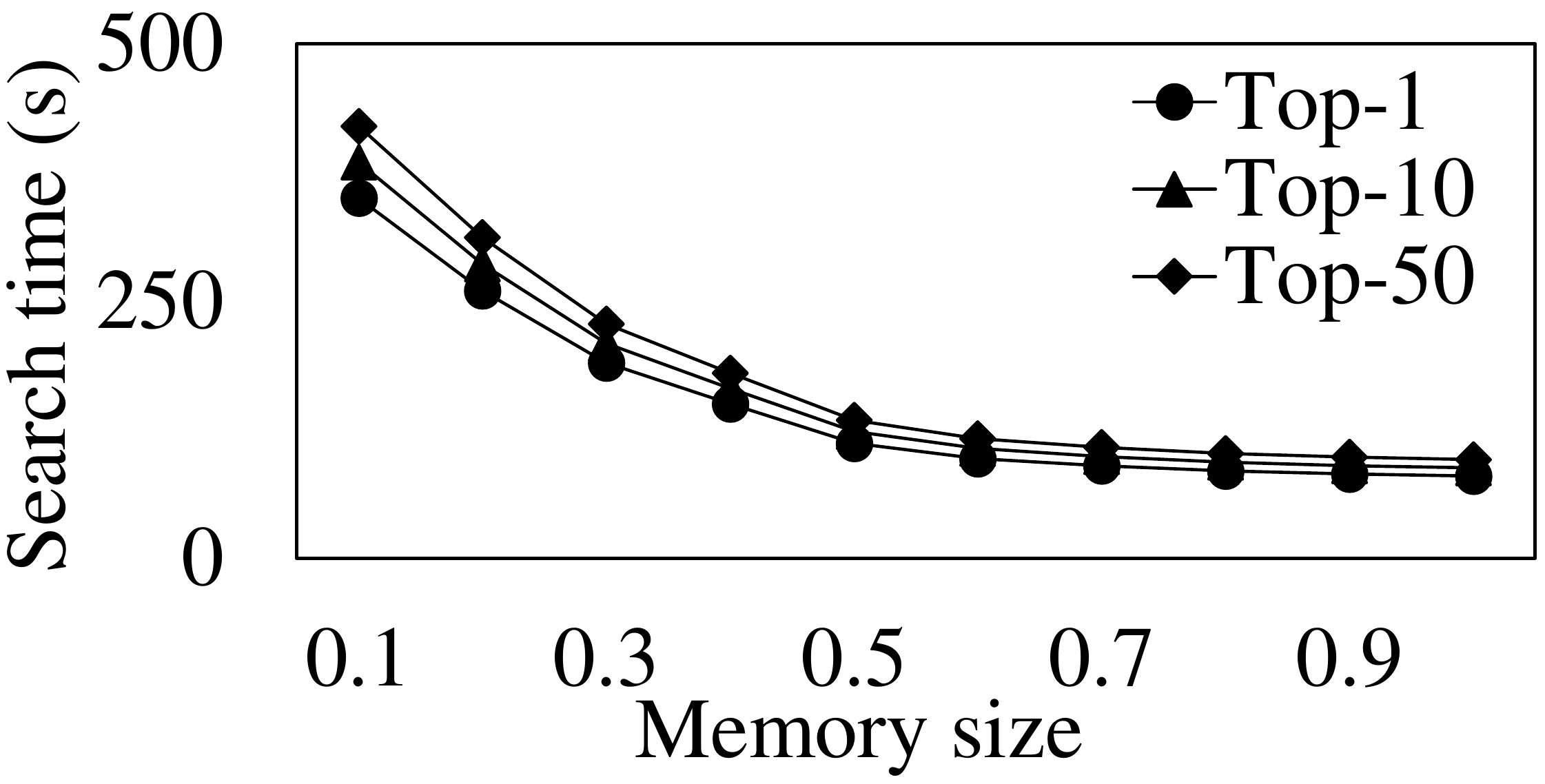}}
   \vspace{-4mm}
   \caption{Search time vs. memory size}
   \label{memory}
   \vspace{2mm}
\end{figure}

If more data can be stored in memory, the time spent to fetch records from disk is  reduced. Therefore, the allocated memory size has an impact on query time. Figure \ref{memory} depicts the time required to answer Top-1, Top-10, and Top-50 queries with 2,000 hash functions under different memory sizes. 

    The horizontal axis in Figure \ref{memory} denotes the allocated memory size (relative size compared to raw data). It is evident that the curves in Figure \ref{memory} depict a descending trend as expected. The curve drops super-linearly with respect to the allocated memory size. The reason is that, the relative position of entities in the MinSigTree is not always guaranteed to be  correlated to their association degrees, especially when the number of hash functions is small (as discussed in Section \ref{sec:parameter}). As a result, although we organize records by their relative position in the MinSigTree, closely associated entities are not always placed in adjacent disk blocks. However, as the memory size reaches $40\%-50\%$ of the dataset size, the curves exhibit only small variation. We also experimented with different measures besides the ADM in Equation (\ref{eq:measure}). Our results indicate that the choice of the measure does not impact the runtime performance of the index and the overall trends remain the same.

\subsection{Sensitivity to result size}
\label{sec:k}
\begin{figure}
   \centering
   \subfloat[REAL data]{\includegraphics[width=.45\columnwidth, height = 2cm]{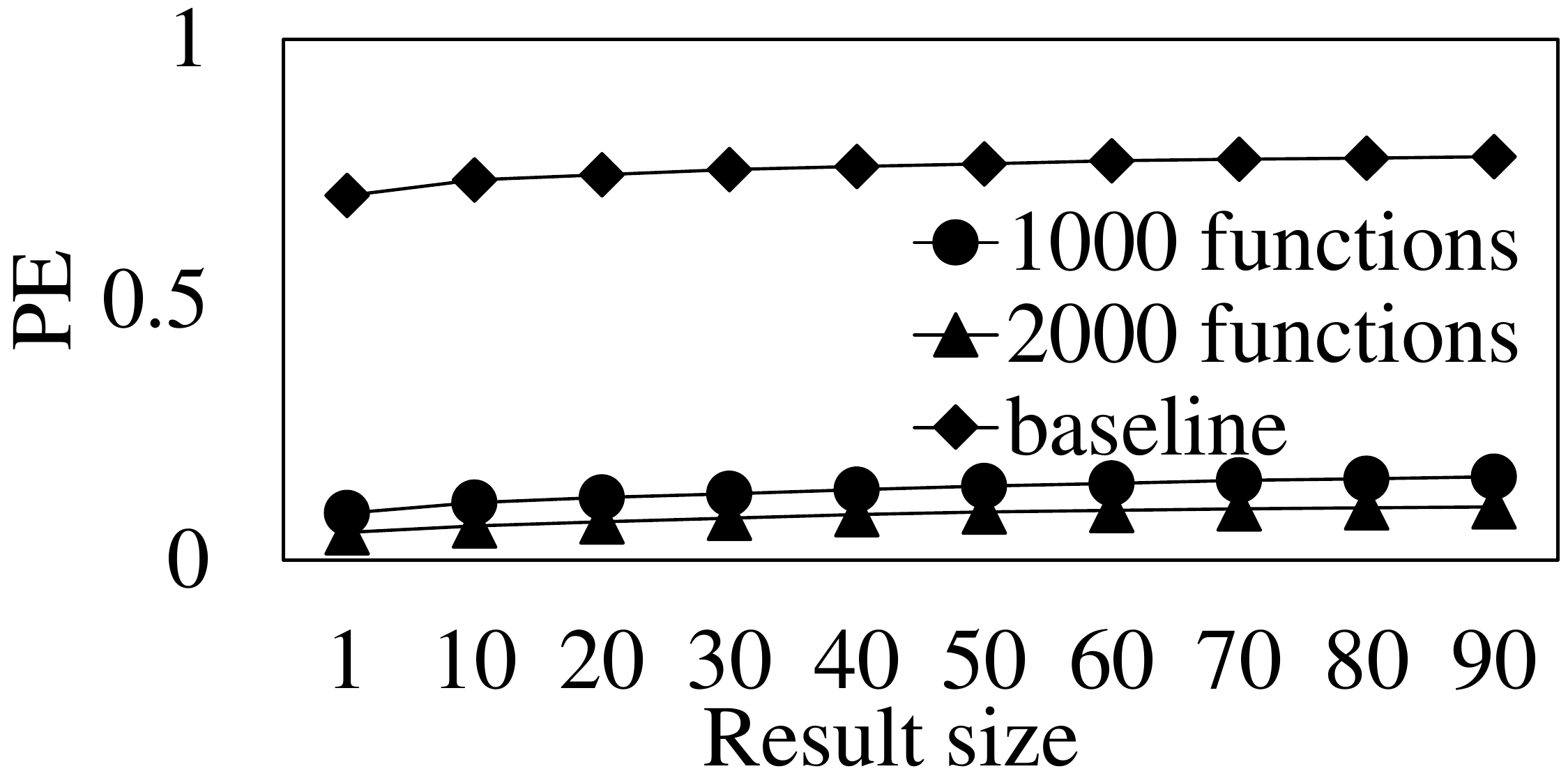}}\quad
   \subfloat[SYN data]{\includegraphics[width=.45\columnwidth, height = 2cm]{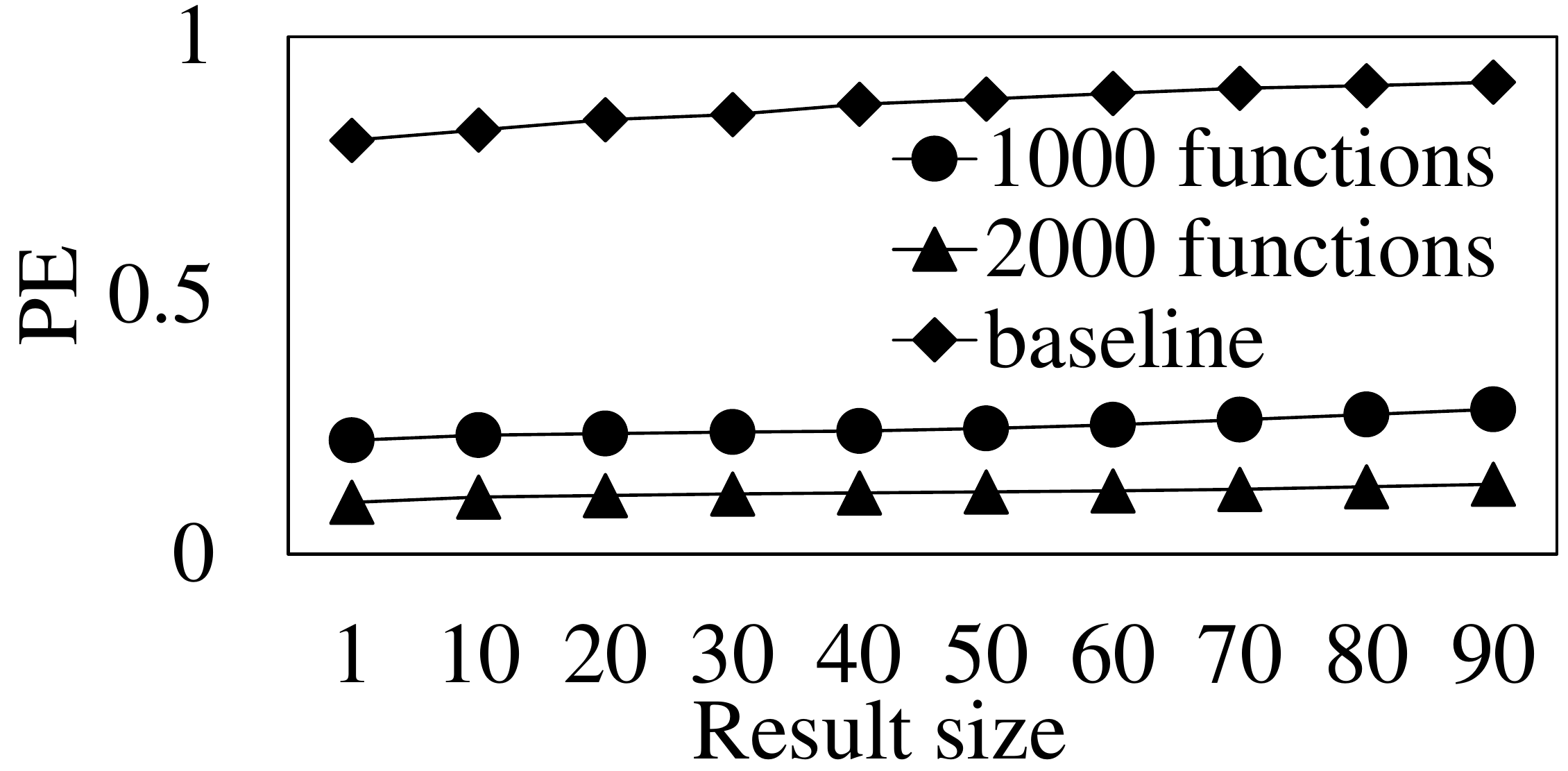}}
   \vspace{-4mm}
   \caption{PE vs. result size ($k$)}
   \label{k}
\end{figure}

We also evaluate our approach as $k$ (number of results desired in top-$k$) increases compared to the baseline method in Figure \ref{k}. PE on both SYN and REAL decreases slightly with increased result size, which is the consequence of both the ADM distribution and the nature of the branch-bound technique. Let $e_q$ be the query entity, $e_a$ be the $i$-th most associated entity to $e_q$, and $e_b$ be the $(i+1)$-th most associated one. Let $df(i)=d(e_q,e_a)-d(e_q,e_b)$ denote the ADM difference between $e_a$ and $e_b$. As Figure \ref{ad} indicates, the association degree distribution ranges for entities are denser when the association degree is small, i.e., $df(i)>df(j)$ if $i<j$ and $df(i)\rightarrow 0$ as $i$ increases. Since the number of hash functions used to compute the signature is far less than the number of ST-cells, the UB of a node is not always guaranteed to be very tight. Let $UB_b$ be the upper bound of the node containing $e_b$, then $d(e_q,e_a)< UB_b$ may occur, especially when $df(i)\approx 0$, i.e., $i$ is large, which means we always need to check $e_b$ before returning $e_a$. As a result, more entities are checked when the value of $k$, i.e., result size, is large, which implies the trend of the curves in Figure \ref{k}.

The baseline method, as argued in Section \ref{sec:baseline}, is based on the existence of clusters among ST-cells, which is not typical in real-life digital traces. Consequently, the PE of the approach is greatly limited, which explains the results in Figure \ref{k} showing that MinSigTree outperforms the baseline approach by large factors.
\subsection{Indexing and update cost}
\label{sec:indexingcost}
\begin{figure}
\centering
\subfloat[]{\includegraphics[width=.45\columnwidth, height = 2cm]{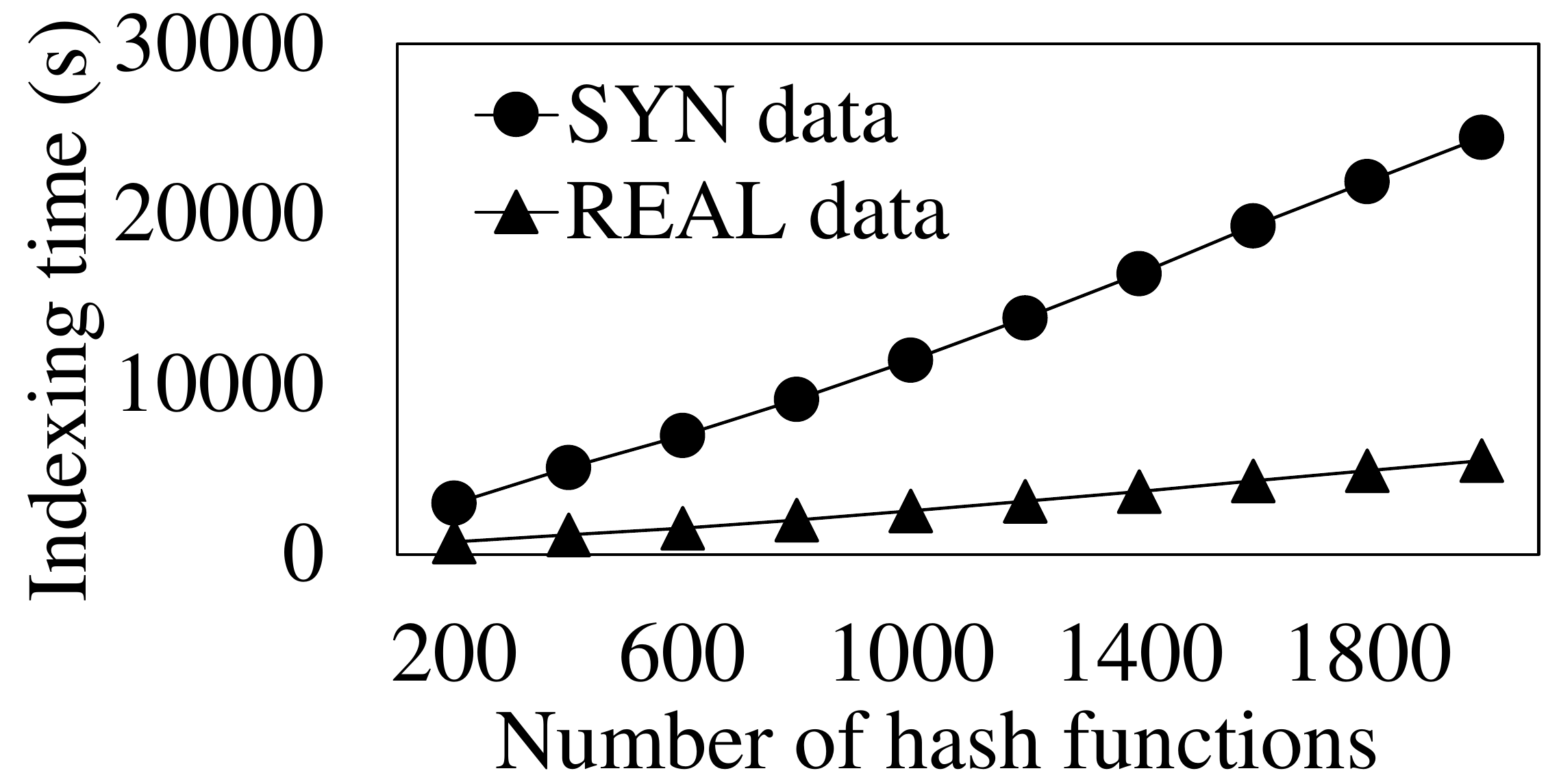}}\quad
   \subfloat[]{\includegraphics[width=.45\columnwidth, height = 2cm]{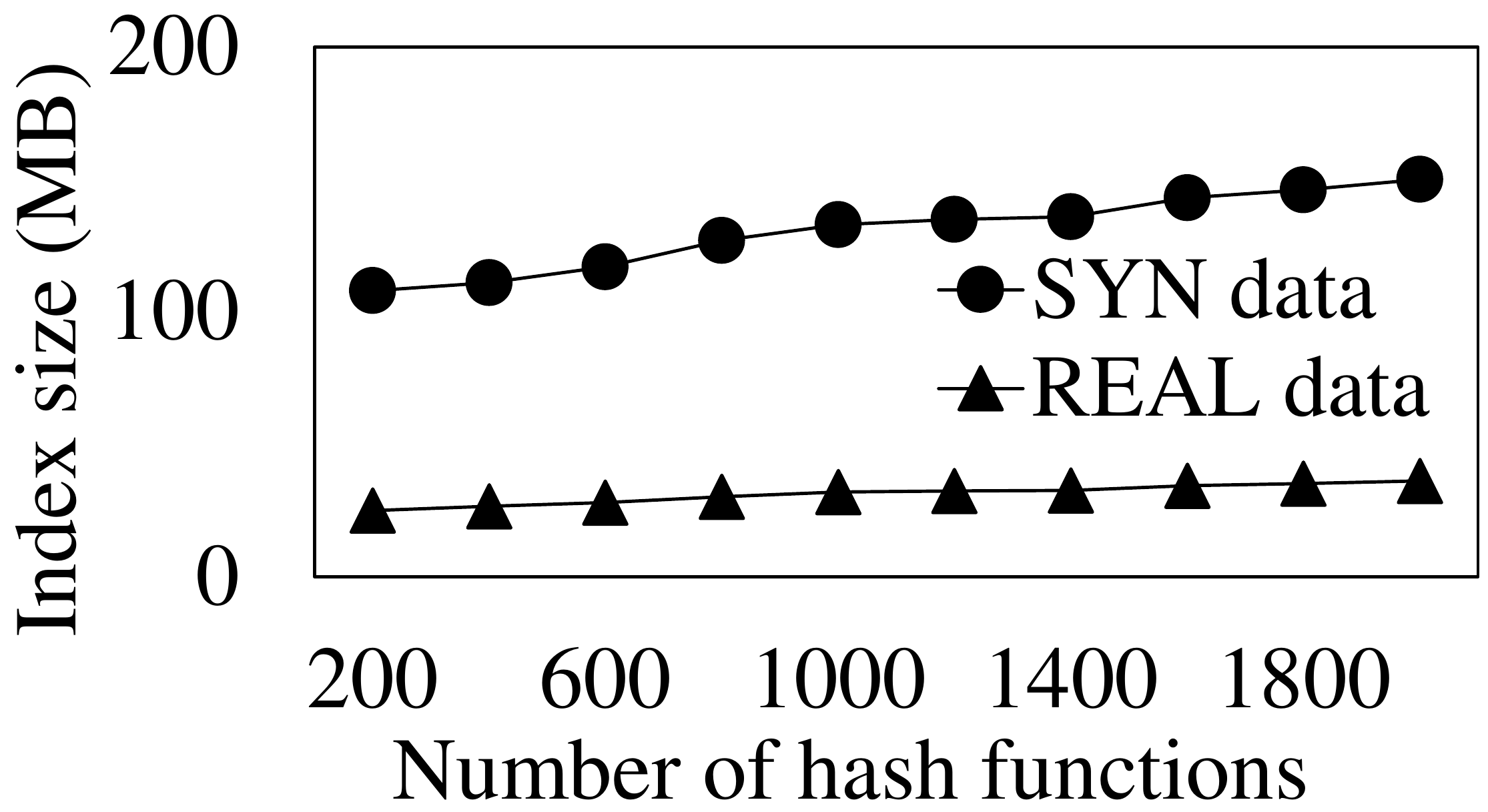}}
   \vspace{-4mm}
\caption{Indexing cost}
\label{build}
\end{figure}

The pre-processing cost to build the MinSigTree is depicted in Figure \ref{build}. Pre-processing time grows almost linearly with the number of hash functions ($n_h$), as the most expensive step in the index construction process is the computation of signatures for each entity, which requires $n_h$ hash operations for each ST-cell where the entity has a presence. 

The size of the MinSigTree is provided in Figure \ref{build}(b). Generally, each node in the MinSigTree contains two integers, one indicating its routing index, and the other recording the hash value of the routing index. A leaf node also includes a pointer to the entities contained in this node. With more hash functions, each entity becomes more unique and thus a node with entity set $\mathcal{E}_N$ may split into several new nodes, each containing a subset of $\mathcal{E}_N$. Therefore, the size of the MinSigTree increases with the number of hash functions. However, the overhead is quite small compared to the data size.
\begin{figure}
\vspace{4mm}
\centering
\includegraphics[width=.45\columnwidth, height = 2cm]{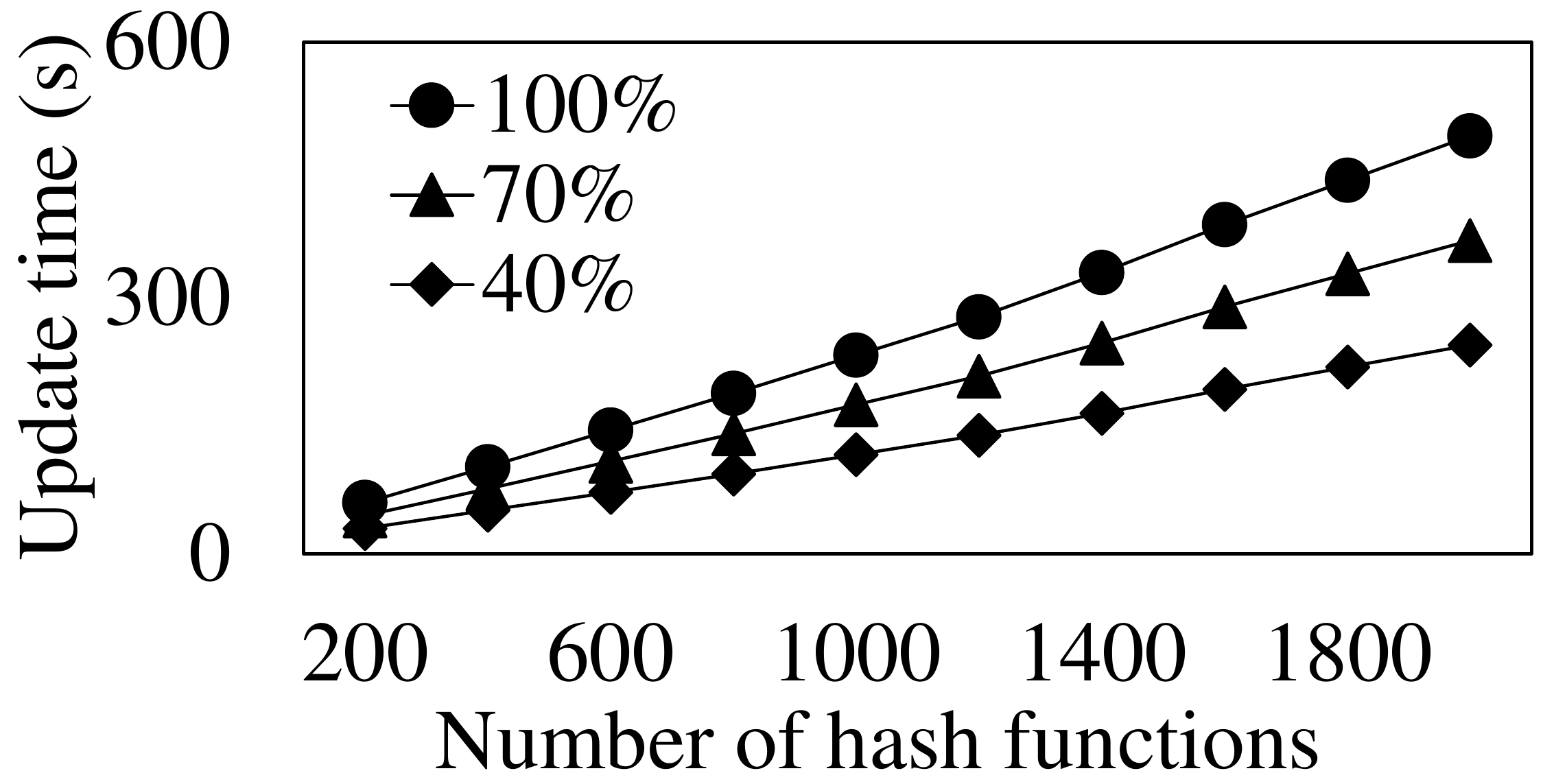}
\vspace{-4mm}
\caption{Update cost}
\label{updatecost}
\end{figure}

Figure \ref{updatecost} illustrates the time required for dynamic index updates. In particular, the figure depicts the time required to update records for 1 million entities in an already built MinSigTree. Since the update process is independent from the data distribution, we present experiments on SYN data. As discussed in Section \ref{sec:term}, we assume that the sp-index remains unchanged over an extended period (as per \cite{wang2016crime}). Thus we focus on two common update operations in practice: inserting new entities and updating existing entities. As presented in Section \ref{sec:incrementalupdate}, when updating existing entities, new digital traces and changes in existing digital traces are processed in the same fashion. Therefore, we do not distinguish between the two cases in this experiment. We report the time required under conditions when $100\%$, $70\%$, and $40\%$ of the entities updated are existing entities, respectively. The time to update grows linearly with the number of hash functions as in the case of building the index. In addition, one can observe that inserting new entities requires less time than modifying the records for existing entities. The reason is that, when updating an existing entity we have to perform the following steps: (1) locate the entity's position in the MinSigTree, (2) remove it from the corresponding leaf node in the index, (3) compute its new signature, and (4) insert it to the proper node. In contrast, for a new entity only steps (3) and (4) are required.

The only parameter that can be tuned by users for performance/cost trade-off is the number of hash functions. In different scenarios one can decide on the suitable number of hash functions utilizing the pruning effectiveness curves of Figure \ref{pruning} and cost curves of Figures \ref{build} and \ref{updatecost}.
\section{Related work}
We are not aware of any work that directly addresses the processing of digital traces as defined herein. Work on query processing over trajectories can be considered related, which focuses on the movement of entities. However the bulk of the work in this area deals with spatial proximity or shape of trajectories and is applicable to moving objects mainly. There are two branches of existing research on querying trajectories, namely, nearest neighbor queries \cite{guting2010efficient,chen2005robust,lee2007trajectory,tang2011retrieving,fang2016scalable,niedermayer2013probabilistic,chen2010searching,lu2011reverse,basu2011location,aly2012spatial,potamias2010k,sharifzadeh2010vor,li2010movemine,abeywickrama2016k,choudhury2016maximizing}, and top-$k$ queries \cite{zheng2015approximate,zheng2013towards,ma2013ksq,wang2017answering,ahmed2017efficient,tang2017extracting,shi2016top,pilourdault2016distributed,shao2016vip,emrich2014extendable,wang2015sharkdb}. Nearest neighbor queries retrieve spatially close trajectories \cite{chen2010searching} or entities \cite{li2010movemine,chen2005robust} under particular distance measures \cite{guting2010efficient,abeywickrama2016k,basu2011location} with various constraints \cite{lee2007trajectory,tang2011retrieving,fang2016scalable,niedermayer2013probabilistic,lu2011reverse,aly2012spatial,potamias2010k,sharifzadeh2010vor,choudhury2016maximizing}, e.g., uncertain trajectories \cite{niedermayer2013probabilistic} or road networks \cite{potamias2010k}. Top-$k$ queries define metrics \cite{ma2013ksq,tang2011retrieving} on specific trajectory types \cite{wang2017answering,ahmed2017efficient,tang2017extracting,shi2016top,pilourdault2016distributed,shao2016vip,emrich2014extendable,zheng2015approximate}, e.g., activity trajectory \cite{zheng2013towards} or semantic trajectory \cite{zheng2015approximate}, to quantify the similarity between trajectories and retrieve trajectories or entities most similar to a given trajectory \cite{ma2013ksq}, entity \cite{zheng2015approximate}, set of locations \cite{zheng2013towards}, etc. One representative piece of work related to the problem studied herein is Frentzos et al. \cite{frentzos2007index}, which proposes a set of metrics for $k$-Most Similar Trajectory ($k$-MST) search over moving object databases, and designs an approximation method for efficient search with R-tree-like structures. These works however along with work on similarity search over trajectories mainly focus on spatial closeness or trajectory shape, without considering the influence of spatial topology, such as hierarchy, on measuring the association among trajectories and the corresponding entities. Consequently they lack the ability to infer the association degree between entities from their trajectories. In addition, as the metrics used therein are based on sequence distance (e.g., Longest Common Sub-Sequence \cite{vlachos2002discovering}), or Time Series distance (e.g., Dynamic Time Warping \cite{sakoe1978dynamic}), which either ignore the time dimension or assume trajectories are aligned in time, they are not guaranteed to satisfy the monotonic properties of the association degree measure of Equation (\ref{eq:d}).

A few pieces of work in recent years also deal with digital traces of human beings \cite{preis2013quantifying,hsieh2016immersive}. Digital traces in their context, however, mainly refer to the records produced by digital devices on the Internet, such as emails, twitter posts, etc, which are not associated with spatial-temporal presences and thus share little semantic similarity with the digital traces proposed in this paper.

Existing top-$k$ query processing techniques, including sorted-list based approaches \cite{fagin2003optimal,fagin2004comparing}, layer based approaches \cite{chang2000onion,xin2006towards}, R-tree based approaches \cite{beckmann1990r,chen2016metric}, are primarily designed to address the problem for low-dimensional data. Since the dimensionality dealt with in this paper is extremely high (as each ST-cell is one dimension, and there are millions of ST-cells), these approaches are not applicable to this problem. On the contrary, the approach proposed in this paper utilizes hashing functions for dimensionality reduction in a fashion that preserves presence instance patterns and thus exact top-$k$ queries can still be supported.

Frequent pattern mining algorithms \cite{aggarwal2014frequent,han2000mining} discover frequently co-occurring items from transaction databases. Such algorithms have also been proposed to identify communities among populations \cite{liu2012mining,feng2016survey}. These approaches are not effective to answer top-$k$ queries in our problem domain as digital traces do not typically contain such patterns.

Hashing techniques are widely adopted in set duplicate detection tasks \cite{song2013inter,gao2014dsh}, among which MinHash demonstrates excellent performance \cite{zheng2016lazylsh,zhu2017interactive,lv2007multi,gan2012locality,satuluri2012bayesian}. Hashing approaches, however, are always used as approximation rather than to calculate exact similarity. We modify MinHash approaches in this paper to support exact top-$k$ queries.
\section{Conclusions and Future Work}
The proliferation of ambient connectivity for certain entity types gives rise to query processing problems of the resulting digital traces. In this paper, we initiated the study and formally defined the problem of top-$k$ query over digital traces, and developed a suite of techniques to efficiently process such queries. We proposed a hash-based indexing structure and combined it with a given spatial hierarchy to answer exact top-$k$ queries, which is a combination that has not been investigated before. We generalized a well-established mobility model to a hierarchical spatial environment and analytically quantified the pruning effectiveness of the proposed method. We also presented extensive experiments on both synthetic and real data sets demonstrating the practical utility of our proposal.

This study introduces several directions for further work. Although top-$k$ query is a natural query to study in this context, several other interesting query processing questions exist. Extending the proposed techniques to other operators such as approximate top-$k$ and joins as well as studying alternate embedding with diverse properties are important directions. Natural extensions to identifying outlier digital traces as well as related data mining questions are worthy of further investigation.
\begin{acks}
This work was supported in part by the NSERC Discovery Grants. We thank
Dr. Parke Godfrey and the anonymous reviewers for their valuable comments and helpful suggestions.
\end{acks}
\bibliographystyle{ACM-Reference-Format}
\bibliography{reference}
\appendix
\section{Notations and Abbreviations}
The notations and abbreviations are given in Table \ref{tab:notation}.
\label{appen:notation}
\begin{table*}[h]
\vspace{10mm}
\caption{Notations and Abbreviations} \label{tab:notation} 
\begin{center}
\begin{tabular}{|c| c|}
\hline
{\bf Notation/Abbreviation} & {\bf Definition}\\

\hline
$\mathcal{E}$ & the set of all entities\\
\hline
$\mathcal{S}$ & the set of all ST-cells\\
\hline
PI & presence instance\\
\hline
$p_a$ & a PI of entity $e_a$\\
\hline
$\mathcal{P}_a$ & the digital traces of entity $e_a$\\
\hline
AjPI & adjoint presence instance\\
\hline
$p_{ab}$ & an AjPI between $e_a$ and $e_b$\\
\hline
$\mathcal{P}_{ab}$ & all AjPIs between entity $e_a$ and entity $e_b$\\
\hline
$seq_a$ & the ST-cell set sequence of entity $e_a$\\
\hline
$sig_a$ & the signature list of entity $e_a$\\
\hline
$\mbox{SIG}_N$ & the signature of node $N$\\
\hline
IM model & individual mobility model\\
\hline
$\alpha$& parameter in IM model controlling the displacement of consecutive PIs\\
\hline
$\beta$& parameter in IM model controlling the duration of PI\\
\hline
$\rho$, $\gamma$ & parameters in IM model controlling the probability of exploratory jump\\
\hline
$\zeta$& parameter in IM model controlling visit frequency\\
\hline
$m$ & level of sp-index\\
\hline
$a$ & width parameter of sp-index\\
\hline
$b$ & relative density parameter of sp-index\\
\hline
$\mathcal{PS}$ & pruned set\\
\hline
$\mathcal{PPS}$ & partial pruned set\\
\hline
PE & pruning effectiveness\\
\hline
ADM & association degree measure\\
\hline
$u$ & parameter in ADM controlling the weight of level\\
\hline
$v$ & parameter in ADM controlling the weight of duration\\
\hline
\end{tabular}
\end{center}
\end{table*}
\section{Cost of index construction}
\label{appen:cost}
\begin{theorem}
\label{th:iocost}
The maximal I/O cost in the indexing process is $2n_p\times \lceil 1+\lceil \log_{n_b} \lceil n_p/n_b\rceil \rceil\rceil + n_p$, where $n_p$ is the total number of pages storing the digital traces and $n_b$ is the number of buffer pages in memory.
\end{theorem}
Methods proposed in Section \ref{sec:approach} require digital traces to be organized by entities, but real world data have varying formats. In a system with adequate memory we can load all records into memory and directly fetch the digital traces of a specific entity. However, when memory becomes the bottleneck, sorting the digital traces by entity becomes a necessity. We employ the well-known $B-$way external merge sort \cite{cole1988parallel} algorithm to do the ordering. Following the sorting principle, the I/O cost in the sorting process is thus $2N\times \lceil 1+\lceil \log_B \lceil N/B\rceil \rceil\rceil$. After the sorting, we need to read the records of each entity once for signature computation.
\begin{theorem}
\label{th:processorcost}
The total processor cost in the indexing process is $\Theta(\vert\mathcal{E}\vert Cmn_h)$, where $C$ is the average number of ST-cells in which an entity has presence.
\end{theorem}
With access to the digital traces organized by entity, we can compute the signature list for each entity and build the MinSigTree. As described in Section \ref{sec:approach}, for each entity, we fetch its $m$ ST-cell sets, employ a family of $n_h$ functions to map each set to a signature, and then build an $m$-level MinSigTree based on the signatures of all entities. Therefore, the total processor cost in the indexing process is $O(\vert\mathcal{E}\vert Cmn_h)$.
\begin{theorem}
\label{th:memoryrequired}
The minimum memory required in the indexing process is $\min \{(n_h)^m, \vert \mathcal{E}\vert\times m\} + n_h + C$.
\end{theorem}
Since the signatures of each entity are computed independently, we can fetch one entity into memory at a time and update the MinSigTree incrementally. In order to avoid extra I/O cost, we need to keep the MinSigTree and hash functions in memory. Theoretically, the size of the MinSigTree is $n_h+(n_h)^2+\cdots +(n_h)^m \approx (n_h)^m$. However, since the total number of entities is $\vert \mathcal{E}\vert$, the number of leaves in the tree is bounded by $\vert \mathcal{E}\vert$. Since each node has one and only one parent node, the number of nodes at other level of the tree is also bounded by $\vert \mathcal{E}\vert$. Therefore, the size of the MinSigTree is $\min \{(n_h)^m, \vert \mathcal{E}\vert\times m\}$. The minimal memory required is thus $(\min \{(n_h)^m, \vert \mathcal{E}\vert\times m\} + n_h + C)$, storing the MinSigTree, $n_h$ hash functions and the ST-cells of one entity. 
\section{Data distribution}
\label{append:datades}
\begin{figure}[h]
   \centering
   \subfloat[REAL data]{\includegraphics[width=.45\columnwidth,height = 2cm]{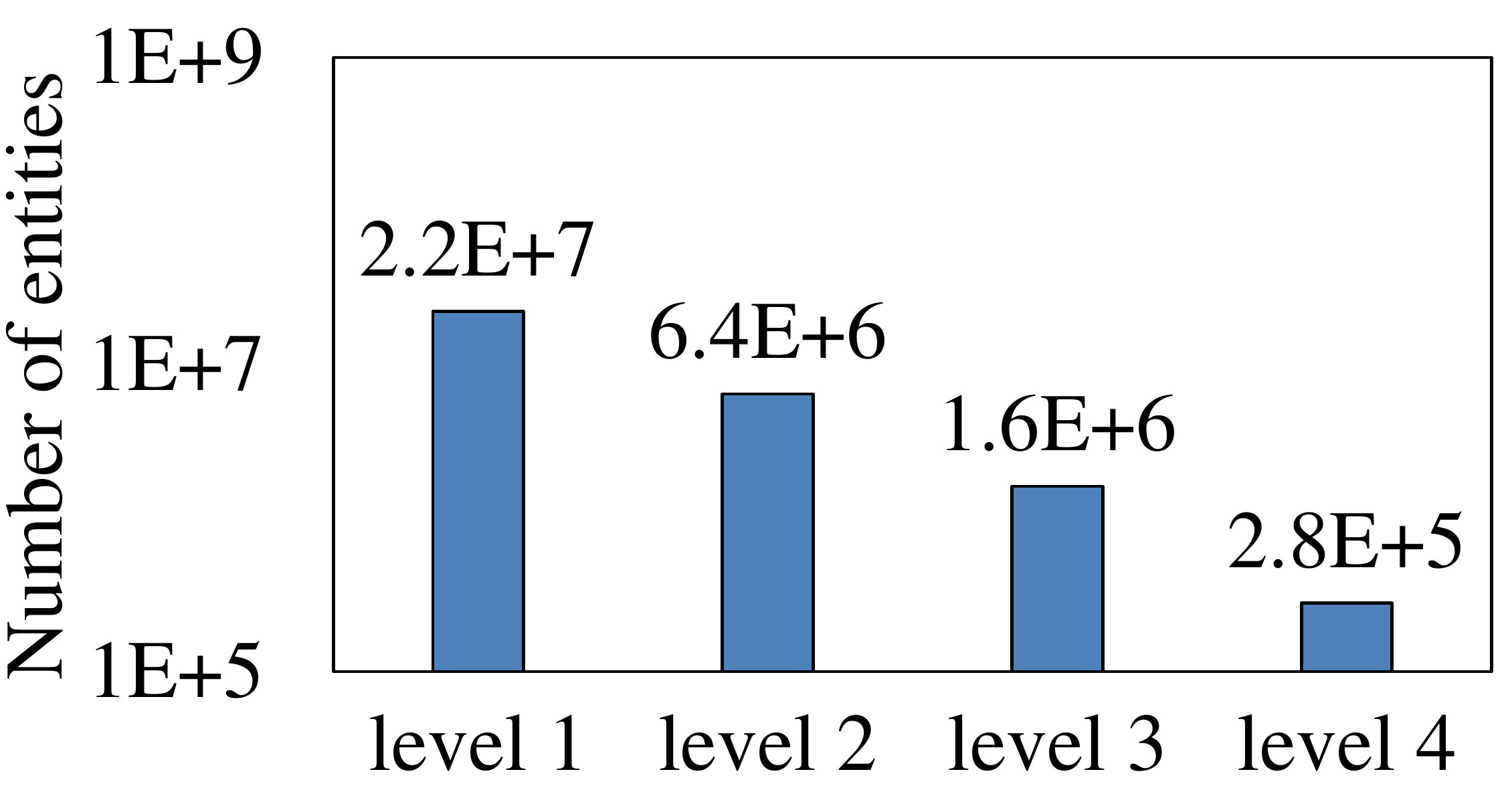}}\quad
   \subfloat[SYN data]{\includegraphics[width=.45\columnwidth,height = 2cm]{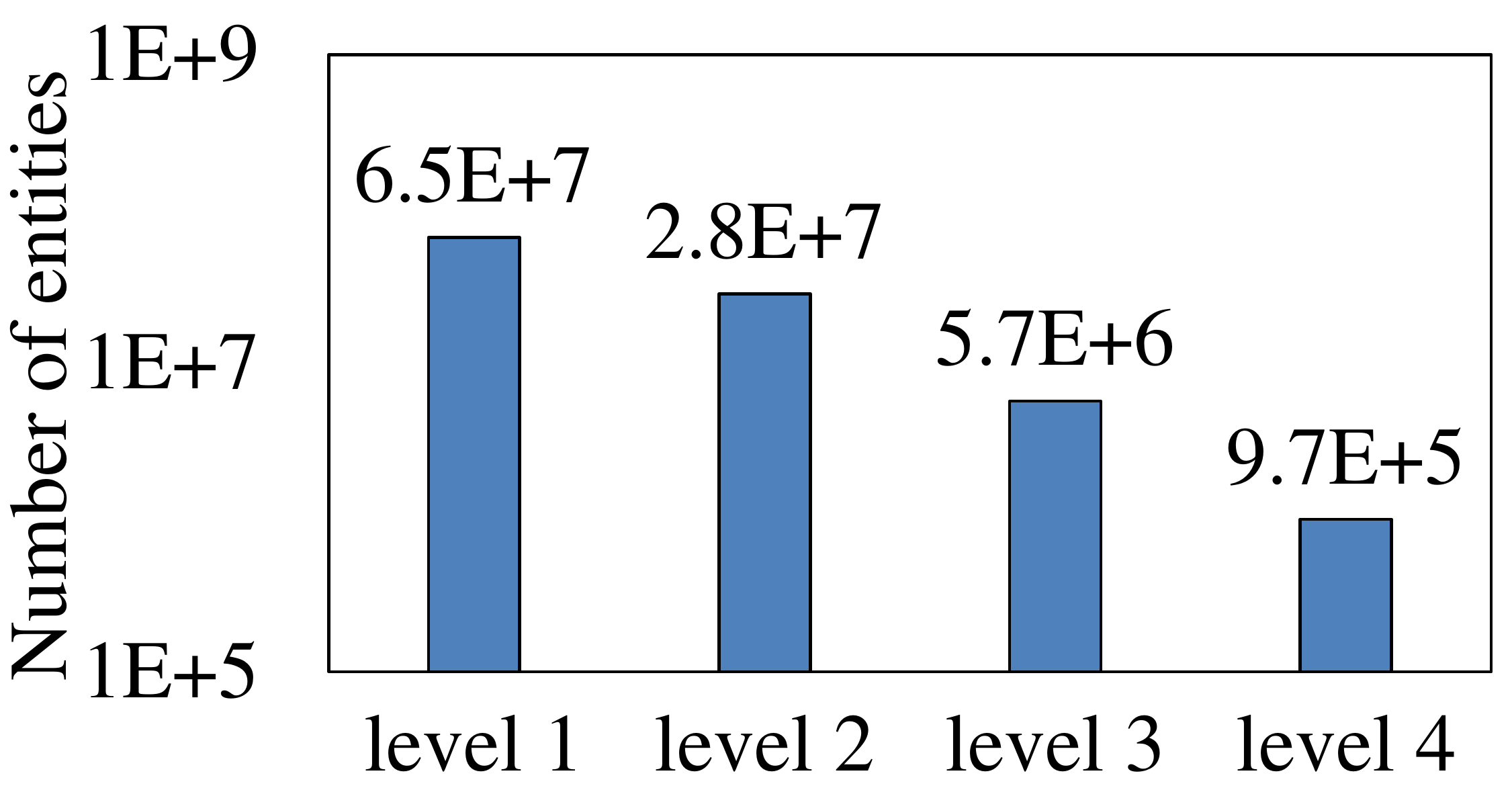}}\\
   \subfloat[REAL data]{\includegraphics[width=.45\columnwidth,height = 2cm]{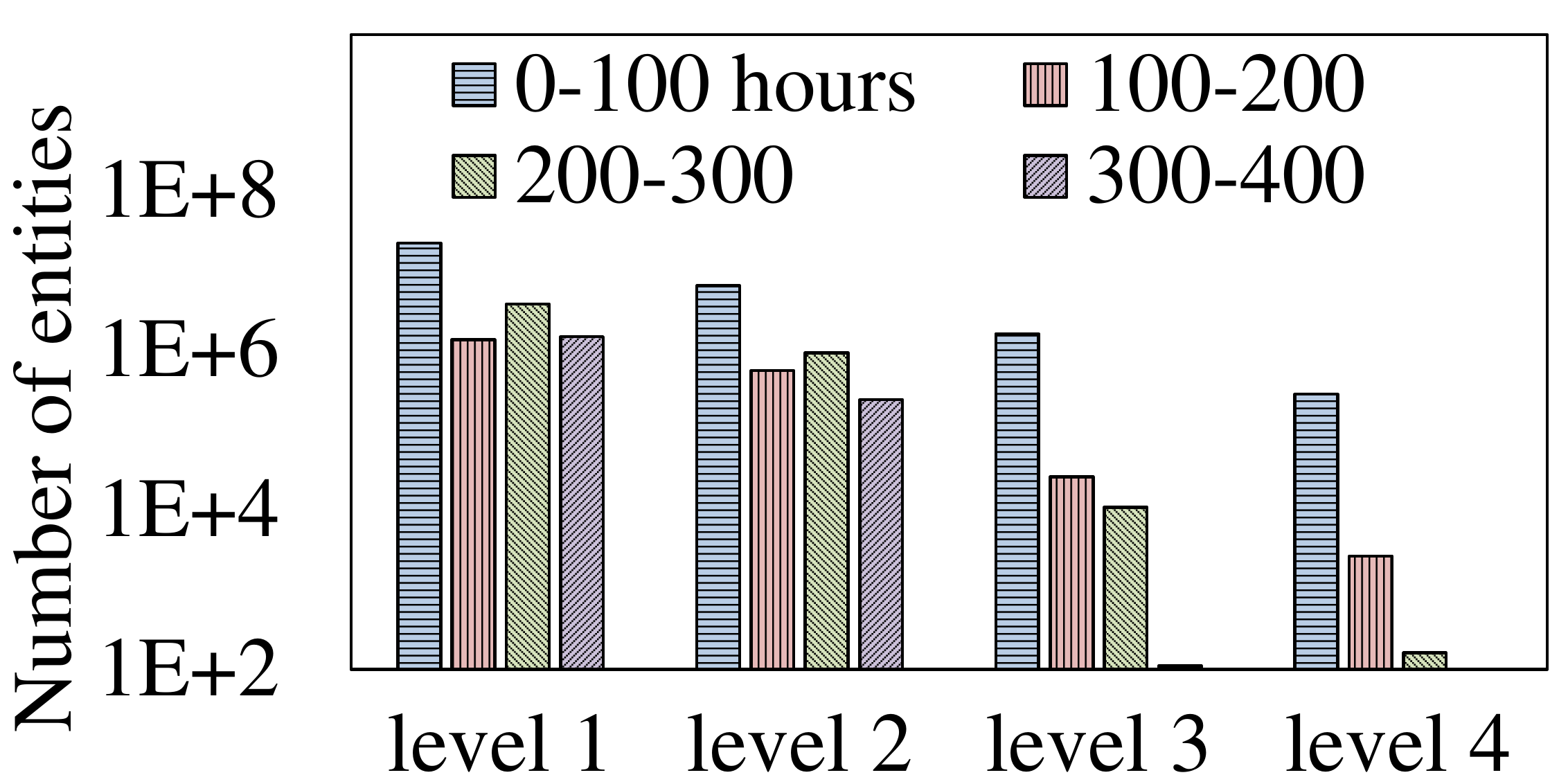}}\quad
   \subfloat[SYN data]{\includegraphics[width=.45\columnwidth,height = 2cm]{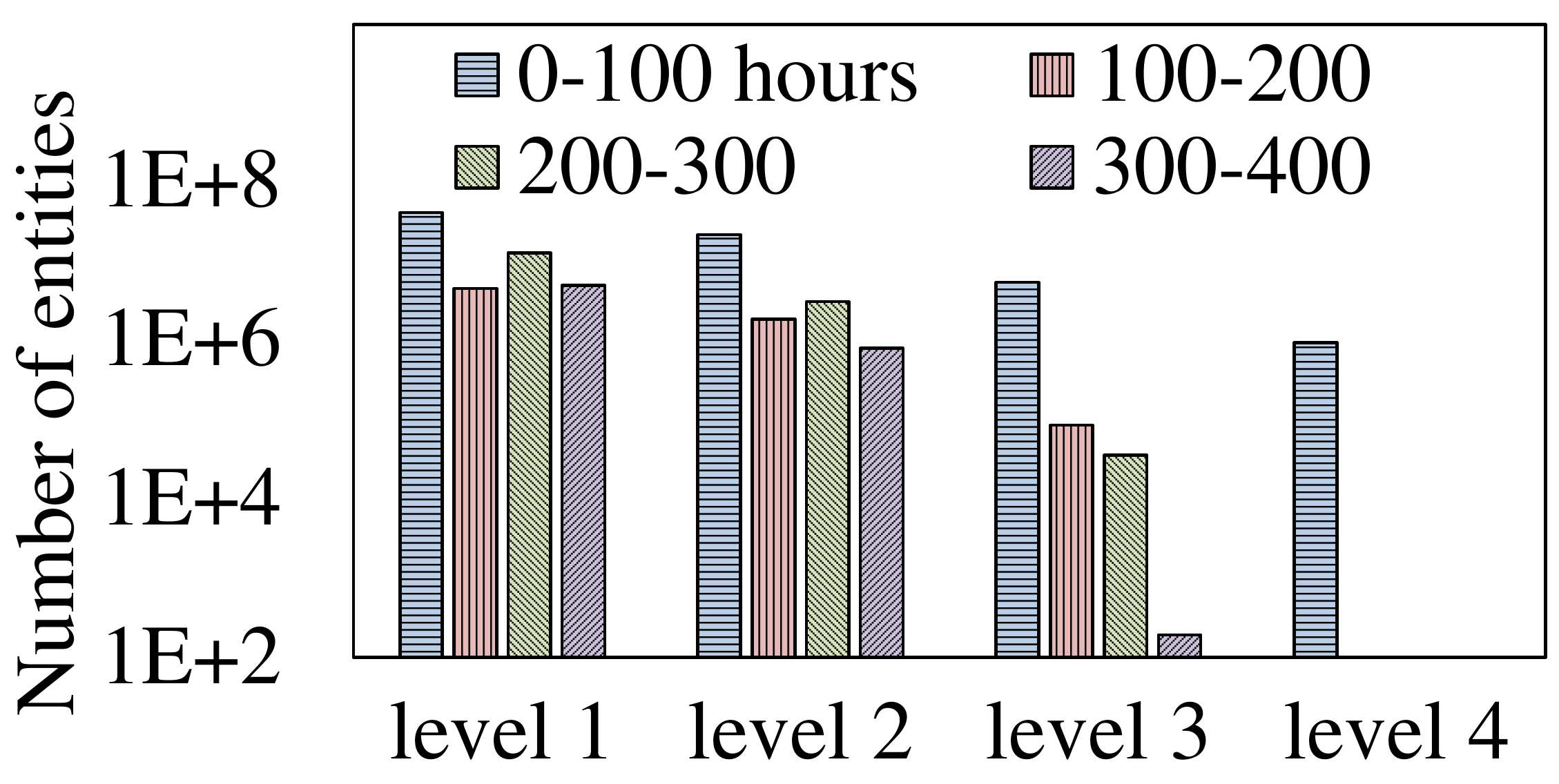}}
   \vspace{-4mm}
   \caption{Data distribution}
   \label{distribution}
\end{figure}

The data distribution is depicted in Figure \ref{distribution}, demonstrating both data distribution across levels as well as distribution of AjPI duration. Note that the vertical axes in all these plots are in log scale. Figure \ref{distribution}(a) depicts the number of entities forming AjPIs with a particular entity at each level on REAL. Given an entity $e$, as shown in Figure \ref{distribution}(a), roughly 22 million entities form AjPIs with $e$ at  level 1 (two entities forming an AjPI at a finer level also form an AjPI at the coarser levels), etc. Figure \ref{distribution}(b) illustrates the same distribution on SYN. Figure \ref{distribution}(c) provides the duration distribution of AjPI at each level: roughly 20 million entities form AjPI with $e$ at level 1 for durations shorter than 100 hours, etc.

\begin{figure}[H]
\centering
\subfloat[REAL data]{\includegraphics[width=.45\columnwidth,height = 2cm]{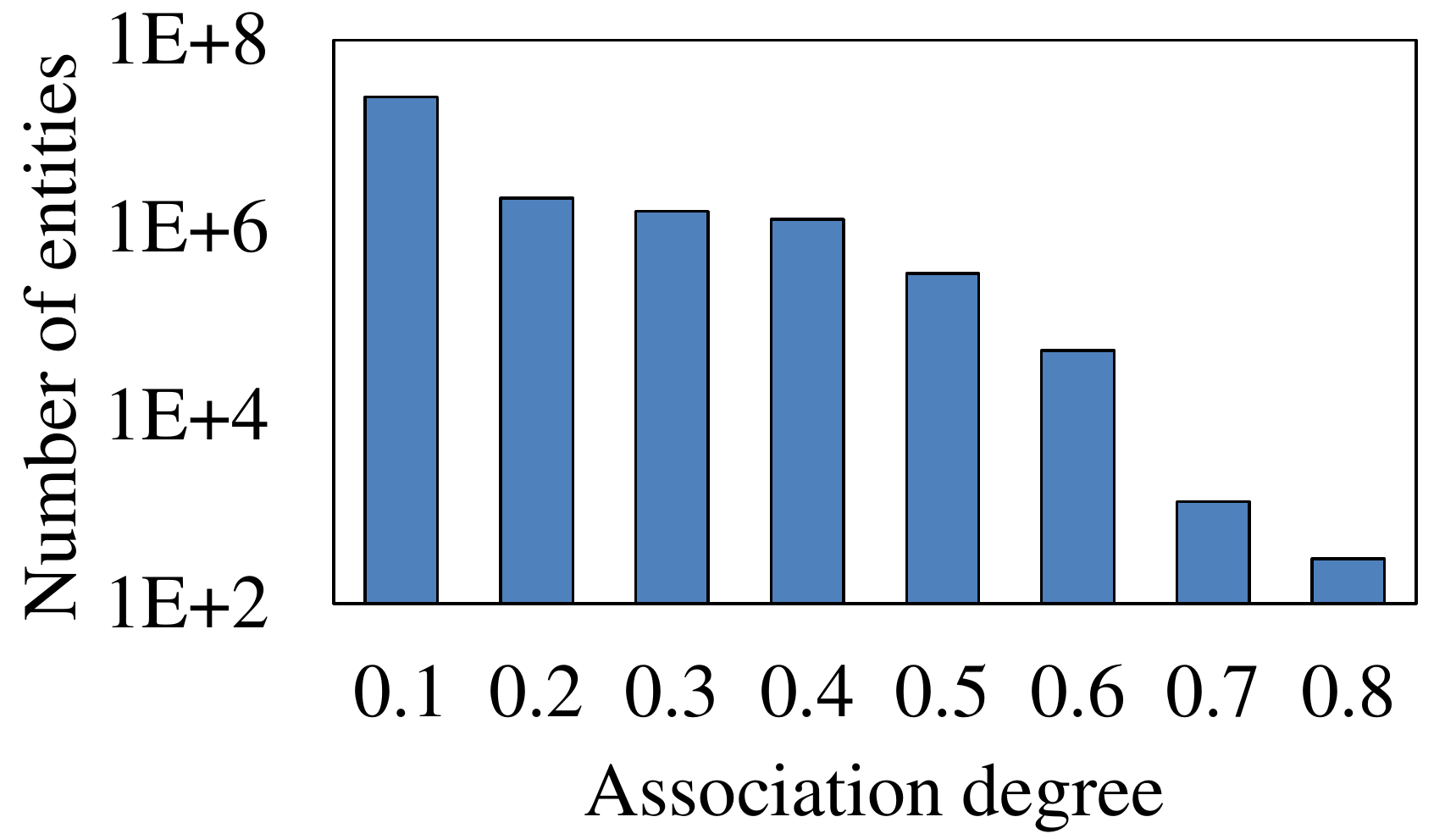}}\quad
   \subfloat[SYN data]{\includegraphics[width=.45\columnwidth,height = 2cm]{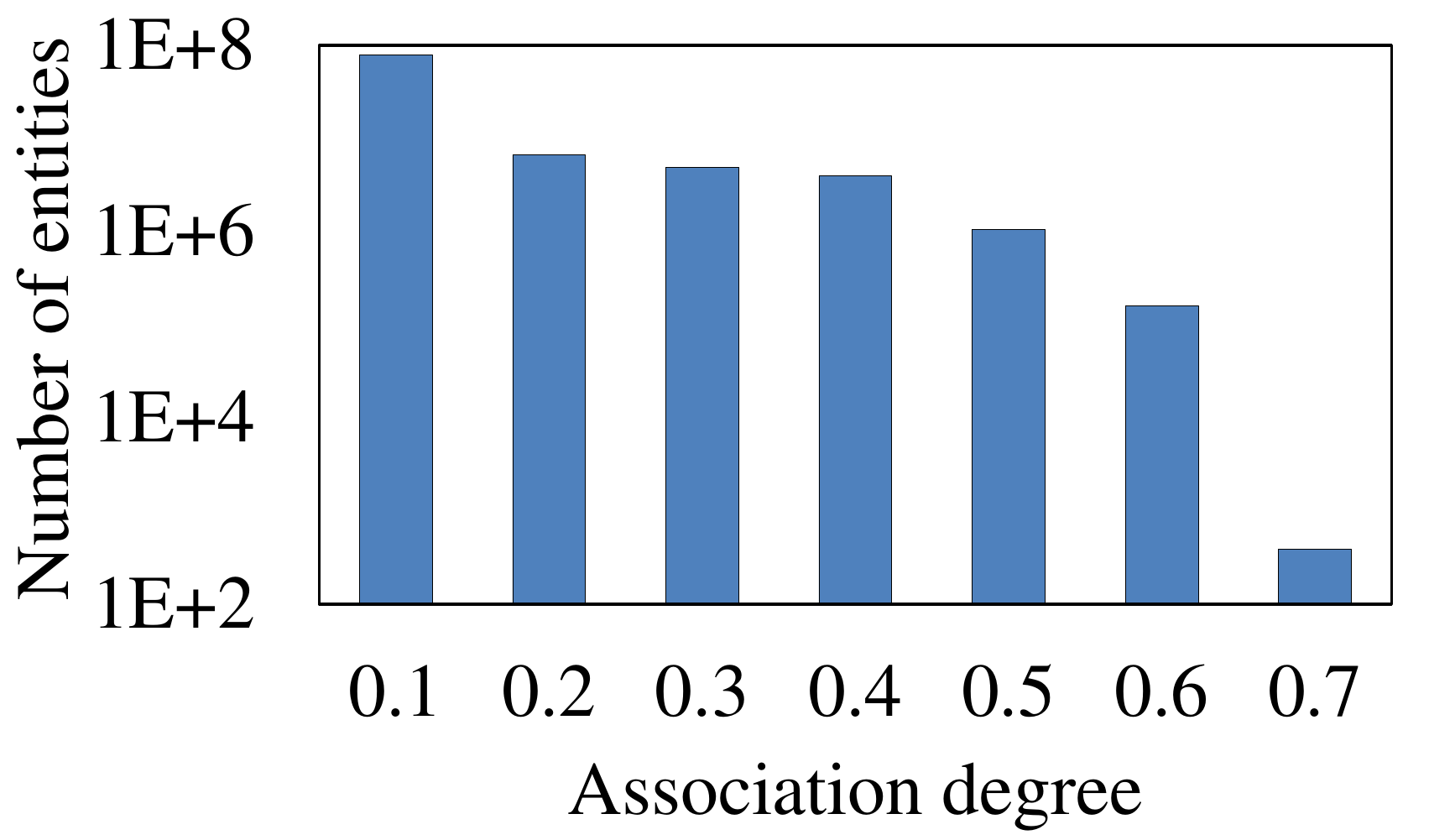}}
   \vspace{-4mm}
\caption{Association degree distribution}
\label{ad}
\vspace{2mm}
\end{figure}

Figure \ref{ad} provides the association degree distribution under $u=1$ and $v=1$, where the horizontal axes are association degree ranges, and the the height of a bar denotes the number of entities falling in the corresponding ADM range with the query entity. From Figure \ref{ad}, it is evident that most entities bear low association degrees with a particular entity.
\section{Measure comparison}
\label{appen:comp}
Intuitively, a similarity measure, $M_p$, \textit{simulates} another similarity measure, $M_q$, on a data set $D$, if the ranking order and the association degrees by $M_p$ and $M_q$ on $D$ are close. Formally, assume that the top-$k$ associated entities to a query entity $e$ measured by $M_p$ form a sequence $R_p$ (sorted by association degree), $R_p^i$ $(i\in [1,k])$ is the $i$-th entity in the ranking; assume that $R_p^i.deg$ denotes the association degree of $R_p^i$ to $e$, then the simulation effectiveness of $M_p$ to $M_q$ is quantified by the metrics in Equation (\ref{eq:coherence}).
\begin{gather}
    K_{avg}(M_p,M_q)=E(K(R_p^\frown \langle \mathcal{R}_q-\mathcal{R}_p\rangle,R_q^\frown \langle \mathcal{R}_p-\mathcal{R}_q\rangle)),
    \nonumber \\
    ADDiff(M_p,M_q)=\frac{\sum_{i=1}^k\vert R_p^i.deg - R_q^i.deg\vert}{k},
    \label{eq:coherence}
\end{gather}
where $K_{avg}$ is a generalized form of Kendall's tau distance \cite{kendall1938new} to measure the distance between top-$k$ lists \cite{fagin2003comparing}, $\mathcal{R}_p$ is the set of entities in $R_p$, $\langle *\rangle$ is an operator transferring a set to a sequence in any order, $^\frown$ denotes the concatenation of two sequences, $E$ is the expectation, and $K$ denotes Kendall's tau distance \cite{kendall1938new} introduced below.

Given two ranked lists, $\tau_1$ and $\tau_2$, both of size $n$, the Kendall's tau distance between $\tau_1$ and $\tau_2$, $K(\tau_1,\tau_2)$, is computed with Equation (\ref{eq:kentau}).
\begin{gather}
    K(\tau_1,\tau_2)=\frac{\vert\{(i.j):i<j,P(i,j)\vee Q(i,j)  \}\vert}{n(n-2)/2},\nonumber\\
    P(i,j)=\tau_1(i)<\tau_1(j)\wedge \tau_2(i)>\tau_2(j),\nonumber\\
    Q(i,j)=\tau_1(i)>\tau_1(j)\wedge \tau_2(i)<\tau_2(j),
    \label{eq:kentau}
\end{gather}
where $\tau_1(i)$ and $\tau_2(i)$ are the ranking orders of element $i$ in lists $\tau_1$ and $\tau_2$ respectively. The Kendall's tau distance between two identical lists is 0, and the Kendall's tau distance between two reverse lists is 1. Kendall's tau distance is commonly used in measuring the ordinal correlation between ranked lists.

Given a data set, $K_{avg}(M_p,M_q)$ describes the consistency of the ranking orders, which corresponds to the effectiveness of the results, and $ADDiff$ depicts the deviation of the association degrees, which influences the performance of the approach. $M_p$ simulates $M_q$ if both $\tau(M_p,M_q)$ and $ADDiff(M_p,M_q)$ are low.
\begin{table}
\vspace{5mm}
\caption{Simulation effectiveness}
\subfloat[][Average Kendall's tau distance]{
\begin{tabular}{c c c c }
        \hline
         & Top-1 & Top-10 & Top-50 \\
         \hline
         Dice&0.0&0.0&0.0 \\
         \hline
         Jaccard&0.0&0.0&0.0 \\
         \hline
         Cosine&2.0E-3&6.7E-3&1.1E-2\\
         \hline
    \end{tabular}}
\\
\subfloat[][Association degree difference]{
\begin{tabular}{c c c c }
        \hline
         & Top-1 & Top-10 & Top-50 \\
         \hline
         Dice&0.0&0.0&0.0 \\
         \hline
         Jaccard&1.1E-2&6.7E-3&5.0E-3 \\
         \hline
         Cosine&3.2E-5&4.0E-5&5.5E-5\\
         \hline
    \end{tabular}}
\label{tab:evaluation}
\end{table}

In order to apply set similarity measures to hierarchical spatial environment, at each spatial level we use Dice, Jaccard, or Cosine metric to compute the similarity between the digital traces of two entities, and use the weighted summation of similarities at all levels as the final association degree. Since weights are independent from the measures and have no influence on the simulation, we simply let the weight of level $i$, $w_i$, take value $\frac{i}{Z}$ ($Z$ is a normalization factor), which corresponds to $u=1$ in Equation (\ref{eq:measure}), and vary the value of $v$ to evaluate the simulation effectiveness of the ADM to other similarity measures.

Table \ref{tab:evaluation} gives the simulation effectiveness of the ADM to other measures. The best simulation to Dice and Cosine Similarity is obtained when $v=1$, and the best simulation to Jaccard Similarity is obtained when $v=1.2$. We can observe from Table \ref{tab:evaluation} that the ADM simulates other measures accurately, especially when the result size ($k$) is small. It is worth noting that the ADM exactly takes the form of Dice Similarity when $v=1$. As is clear from Equation (\ref{eq:measure}), varying the value of $v$ only changes the association degree, but has no influence on the ranking order. 
Our experiments indicate that when $v$ is in the range of $[0.5,2]$ the association degree computed by the ADM is close to those of the other measures, and thus we utilize values in this range during experiments.

\end{document}